\documentclass[12pt, draftclsnofoot, onecolumn]{IEEEtran}
\usepackage[utf8]{inputenc}
\makeatletter
\def\ps@headings{%
\def\@oddhead{\mbox{}\scriptsize\rightmark \hfil \thepage}%
\def\@evenhead{\scriptsize\thepage \hfil \leftmark\mbox{}}%
\def\@oddfoot{}%
\def\@evenfoot{}}
\makeatother 
\usepackage{amsfonts,dsfont}
\usepackage[dvips]{graphicx}
\usepackage{caption}
\usepackage{subfigure}
\usepackage{times}
\usepackage{cite}
\usepackage{lettrine}
\usepackage{amsmath}
\usepackage{bm}
\allowdisplaybreaks[4]
\usepackage{array}
\usepackage{amssymb}

\usepackage{stfloats}
\usepackage{slashbox}
\usepackage{graphicx}
\usepackage{footnote}
\usepackage{booktabs}
\usepackage{array}
\usepackage{algorithmic}
\usepackage{algorithm}
\usepackage{subeqnarray}
\usepackage{cases}
\usepackage{threeparttable}
\usepackage{color}
\usepackage{url}

\usepackage{cleveref} 

\makeatletter       
\renewcommand{\maketag@@@}[1]{\hbox{\m@th\normalsize\normalfont#1}}%
\makeatother
\makeatletter
\renewcommand{\fnum@figure}{Fig. \thefigure}
\makeatother

\DeclareMathOperator*{\argmax}{argmax}

\newtheorem{theorem}{\underline{Theorem}}
\newtheorem{lemma}{\underline{Lemma}}

\newtheorem{proposition}{\underline{Proposition}}
\newtheorem{remark}{\underline{Remark}}

\begin{document}
\title{Multi-Cell Mobile Edge Computing: Joint Service
Migration and Resource Allocation}
\author{Zezu Liang, Yuan Liu, Tat-Ming Lok, and Kaibin Huang

\thanks{Z. Liang and T. M. Lok are with Department of Information Engineering, The Chinese University of Hong Kong, Hong Kong (e-mail: lz017@ie.cuhk.edu.hk;  tmlok@ie.cuhk.edu.hk). Y. Liu is with school of Electronic and Information Engineering,
South China University of Technology, Guangzhou 510641, China (e-mail: eeyliu@scut.edu.cn).  K. Huang is with Department of Electrical and Electronic Engineering, The University of Hong Kong, Hong Kong (e-mail: huangkb@eee.hku.hk).}
}

\maketitle

\vspace{-1.5cm}
\begin{abstract}
Mobile-edge computing (MEC) enhances the capacities and features of mobile devices by offloading computation-intensive tasks over wireless networks to edge servers. One challenge faced by the deployment of MEC in cellular networks is to support user mobility. As a result, offloaded tasks can be seamlessly migrated between base stations (BSs) without compromising the resource-utilization efficiency and link reliability. In this paper, we tackle the challenge by optimizing the policy for migration/handover between BSs by jointly managing computation-and-radio resources. The objectives are twofold: maximizing the sum offloading rate, quantifying MEC throughput, and minimizing the migration cost. The policy design is formulated as a decision-optimization problem that accounts for virtualization, I/O interference between virtual machines (VMs), and wireless multi-access. To solve the complex combinatorial problem, we develop an efficient relaxation-and-rounding based solution approach. The approach relies on an optimal iterative algorithm for solving the integer-relaxed problem and a novel integer-recovery design. The latter outperforms the traditional rounding method by exploiting the derived problem properties and applying matching theory. In addition, we also consider the design for a special case of ``hotspot mitigation", referring to alleviating an overloaded server/BS by migrating its load to the nearby idle servers/BSs. From simulation results, we observed close-to-optimal performance of the proposed migration policies under various settings. This demonstrates their efficiency in computation-and-radio resource management for joint service migration and BS handover in multi-cell MEC networks.
\end{abstract}


\section{Introduction}
Mobile (or multi-access) edge computing (MEC), which provides users computing services at the network edge, is envisioned as a key technology in the fifth generation (5G) systems for supporting computation-intensive and latency-critical mobile applications \cite{White_Paper,survey}. In MEC systems, the computation intensive tasks of mobile users are offloaded to edge servers co-located with base stations (BSs) or access points. This avoids data transportation to the remote cloud centers, thereby dramatically reduce latency and avoid traffic congestion in the backhaul network. In this work, we address the issue of supporting mobility in an MEC network, referring to a cellular network providing MEC services. In a traditional radio access network, a key solution for mobility is to handover a travelling user's wireless link from one BS to another to ensure its reliability \cite{BS_handover}. The handover in an MEC network is more complex as it may also involve the migration of computing tasks between servers, called \emph{service migration} \cite{VMmigration}. Making a migration decision should account for factors including computation resources and load at two servers/BSs and the migration cost incurred by data transportation across the backhaul network. To tackle the challenges, we propose in this paper a framework of \emph{joint migration-and-handover (JMH)} for multi-user multi-cell MEC systems.

\vspace{-0.15in}
\subsection{Resource Management in MEC Networks}

Among others, one vein of MEC research that is aligned with the current work is resource management. It features the joint management of computation and radio resources to achieve a high efficiency for computation offloading. A binary offloading policy is proposed in \cite{singleuser1} for adapting the offloading decision to a stochastic wireless channel under the criterion of minimizing mobile energy consumption. Peer-to-peer cooperative MEC is proposed in \cite{NOMA} where one mobile device serves as a helper for another by offloading the latter's computation or relaying it to a server. For multi-user MEC systems, the resource management is more complicated since it involves resource sharing by multiple offloading users. In \cite{Multiuser1}, a centralized resource allocation scheme is proposed for minimizing sum mobile energy consumption. The design is extended in \cite{Asynchronous} to the case of asynchronous offloading. On the other hand, distributed resource allocation strategies can be designed by applying game theory, which is pursued in \cite{multiuser2,multiuser3}. In a multi-user and multi-server system, there is an additional issue of load distribution among servers. It is addressed in \cite{bao} by an efficient distributed offloading design based on matching theory and in \cite{RL} using the reinforcement-learning approach. It has also been studied in various MEC system configurations like vehicular networks \cite{multiserver3} and unmanned aerial vehicle (UAV) systems \cite{UAV2}.

Another important type of resources in computing is I/O resources such as the bandwidth of a bus connecting a GPU and its associated system. For edge or cloud computing based on virtualization, tasks are executed simultaneously in the same server in the forms of virtual machines (VMs). The sharing of finite I/O resources by VMs causes mutual computing interference, called I/O interference, which slows down their computing speeds \cite{W,netIO,VM1}. Being a potential performance bottleneck, I/O interference is extensively studied in the area of cloud computing to understand its effects and find solutions (see e.g., \cite{VM1}). In contrast, I/O interference is not yet extensively studied in the area of MEC despite some recent work on factoring it into the design of offloading policies \cite{our_work}. In this work, we also consider I/O interference in JMH.

In view of prior work, existing results on resource management for MEC focus on the optimization of offloading policies. In this work, we explore an uncharted direction of resource management for migration and handover to support mobility in MEC networks. Most existing work focuses on sharing the resources of a single server (or server cluster) by multiple offloading users. In contrast, we focus on the balancing of the resources among servers/BSs by controlling both migration and handover.

\vspace{-0.15in}
\subsection{Service Migration and BS Handover}

As a key mechanism for dynamic resource management, service migration has been widely studied in the area of cloud computing covering a wide range of topics including network load balancing \cite{load_balance}, hotspot mitigation \cite{hot_spot}, and I/O interference aware migration \cite{Migration_policy}. Migration in cloud computing targets a wired network (e.g., server grid within a data center) where links are assumed reliable. In contrast, the implementation of service migration in an MEC network will be inevitably coupled with the handover of wireless links over BSs. The links experience fading and each BS serves a dynamic number of users and hence has time-varying available radio sources besides a random computation load. The coupling between service migration and BS handover calls for their joint design to improve the offloading performance of the MEC network, which forms the theme of this work.

In traditional wireless networks, BS handover is incurred by deterioration in wireless link quality of the serving BS and is employed to re-associate with another for higher radio access. However, such handover mechanisms are not sufficient to support efficient computation offloading in an MEC network. On one hand, as mentioned earlier, handover of MEC services is conducted in JMH for ensuring radio and computing reliability, and thereby the migration cost on both sides should be taken into account. On the other hand, apart from channel condition, the computation capabilities of different BSs need to be considered when associating a user with an appropriate BS.
The work \cite{handover_timing,multiceil2,dynamic,followme} investigates BS handover in MEC under mobility consideration, which however does not consider the variation of computation resource by BS handover. In contrast, our proposed JMH framework considers that the computing speeds of two servers/BSs fluctuate caused by handover. Such an issue is not studied yet in MEC migration/handover.

\vspace{-0.15in}
\subsection{Our Contributions}
In this paper, we study the problem of optimal JMH in a multi-user multi-cell MEC system based on virtualization. The optimization problem aims at maximizing the weighted sum offloading rates of all the users while minimizing the incurred migration cost as much as possible by controlling the migration/handover decisions. The problem accounts for both I/O interference and multi-user interference.

The main contribution of the work lies in developing a practical algorithm for designing the optimal JMH policy. The said problem is an integer nonlinear program and nonconvex. To overcome the difficulty, we propose a two-stage solution method. First, the binary constraints of the migration decisions are relaxed, allowing fractional programming to be applied to solve the relaxed problem. Next, a novel rounding method based on the problem's properties is proposed for recovering the binary decision solution, which outperforms the naive rounding method.

The other contribution of the work is to optimize JMH for the hotspot mitigation scenario, referring to alleviating an overloaded server by migrating its load to the helper servers. We show that when the load of the hotspot server is below a certain threshold, the optimal JMH scheme can effectively address the overloaded condition of the hotspot server via load balance among servers. When the load exceeds the threshold, all the servers are overloaded after optimal JMH and in this case adding more helper servers is needed.

The rest of this paper is organized as follows. In Section II, we present the system model and problem formulation. We introduce the algorithm to solve the formulated problem in Section III and discuss the special case of the hotspot mitigation in Section IV. Finally, simulation results and conclusions are provided in Section V and Section VI, respectively.

\begin{figure}[t]
\setlength{\abovecaptionskip}{0.4cm}
\setlength{\belowcaptionskip}{-0.4cm}
\begin{centering}
\includegraphics[width=0.55 \linewidth]{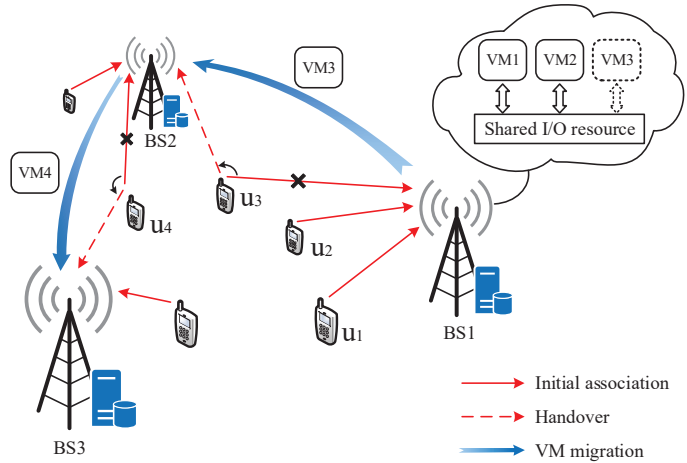}
\vspace{-0.1in}
\caption{A multi-cell MEC system, where $u_3$ and $u_4$'s services are enabled by joint BS handover and VM migration.}\label{fig:1}
\end{centering}
\end{figure}

\vspace{-0.1in}
\section{System Model and Problem Formulation}\label{sec2}
\vspace{-0.1in}
\subsection{System Model}
As shown in Fig. \ref{fig:1}, we consider an MEC system consisting of $N$ BSs and $K$ users, denoted by the set $\mathcal{N}=\{1, \cdots, N\}$ and set $\mathcal{K}=\{1, \cdots, K\}$, respectively. Each BS is integrated with a server that can provide computing services to users if it hosts the users' corresponding VMs. We assume that each user is served by one dedicated VM and each VM only serves the corresponding user\footnote{An VM dedicated to serving a user is commonly used for the user's service with the requirements in user-customized compute environment, data security and confidentiality, and/or performance stability. For example, each Google Docs service gets an VM/container to serve a specific user.}. A VM is a software clone of user's service environment, which contains the user's profiles and applications for running user's offloaded tasks and can be migrated among BSs to continue serving the user wherever it moves. In the proposed JMH framework, when a user switches its association from one BS to another, the user's corresponding VM is also migrated between the two BSs (e.g., see users $u_3$ and $u_4$ in Fig. \ref{fig:1}). We assume a time-slotted model that the user-BS associations and channel gains remain unchanged in each slot but can be varied from one slot to another. The channel gains are considered to include path loss and shadowing while neglect the small-scale (fast) fading, in view of the fact that small-scale fading has little effect on reference signal receiving quality (RSRQ) that is the measurement for handover and dominated by path loss in practice. Moreover, the effect of small-scale fading can be averaged out by employing a sufficiently long channel code in practice \cite{multi_server,UAV}. Thus, the channel gains can be regarded as static within each slot but may vary from one slot to another. Let $x_{k,n}$ denote the JMH decision for the service migration, with $x_{k,n}=1$ indicating the service of user $k$ is placed at BS $n$ and $x_{k,n}=0$ otherwise. We assume that each user can associate with only one BS, thus, $\sum_{n\in\mathcal{N}}x_{k,n}=1$, $\forall k\in\mathcal{K}$. The JMH process can incur system overheads, such as consuming backhaul bandwidth to transfer VM data and the handover signaling. To account for this, we consider a fixed cost $c_{k,j,n}$ occurs when user's $k$ service is migrated from BS $j$ to BS $n$, with $n\neq j$. We assume $c_{k,j,n}=0$ if $n=j$ (i.e., no cost occurs if not migrated). Then, given the current service locations $\{x_{k,n}^{0}\}$, the JMH cost of each user in next time slot is given by
\begin{align}
C_k=\sum_{n\in\mathcal{N}}\sum_{j\in\mathcal{N}}x_{k,j}^{0} x_{k,n}c_{k,j,n},
\end{align}
where $x_{k,j}^0x_{k,n}$ indicates whether user $k$'s service is initially placed on BS $j$ and to be migrated to BS $n$ (i.e., $x_{k,j}^0x_{k,n}=1$) or not (i.e., $x_{k,j}^0x_{k,n}=0$). For simplicity, we assume that the migration/handover time is negligible compared with the slot length.

\subsubsection{Communication Model}
Denote the uplink channel gain from user $k$ to BS $n$ as $g_{k,n}$, the transmit power of user $k$ as $p_k$, and the noise power of BS $n$ as $\sigma_n^2$. For the ease of problem analysis, we consider the simple case that users offload data at the same band, i.e., frequency reuse factor of $1$. The extension to radio resource allocation among users will be elaborated in Section \ref{ssec:BA}. Then, the achievable uplink transmission rate for user $k$ offloading tasks to BS $n$, denoted as $r_{k,n}$, is
\begin{align}\label{eqn:1}
r_{k,n}=B\log_2\left(1+\frac{p_kg_{k,n}}{\sigma_n^2+\sum_{j\in\mathcal{K}\backslash\{k\}}p_jg_{j,n}}\right).
\end{align}
where $B$ denotes the system bandwidth. As the transmit powers are assumed to be fixed, each user's transmission rate $r_{k,n}$ is deterministic through \eqref{eqn:1}. We ignore the result downloading phase because of the relative much smaller sizes of computation results \cite{multiceil2}.

\subsubsection{Computation Model}
The MEC server at each BS accommodates users' offloaded tasks into their own VMs and executes them in parallel, namely \emph{parallel computing}. We consider the I/O interference in parallel computing \cite{W,netIO} and adopt a model developed in the literature \cite{parallel_computing_mode} to characterize the computation rate. Specifically, let $f_{k,n}$ denote the expected computation rate (offloaded bits per second) of user $k$'s VM when running in isolation at BS $n$. Following \cite{parallel_computing_mode}, we define $d_n>0$ as a performance degradation factor\footnote{To be more specific, $d$ is defined as the percentage increase in the computing time (i.e., $1/f$) of a VM when multiplexed with another VM on the same server. We assume the server can coordinate the workloads and resource demands of VMs to reach a homogenous factor $d$ \cite{parallel_computing_mode}. Then, indicating with $T$ the computing time of a VM in isolation, the expected computing time of a VM when multiplexed with an additional VM can be expressed as $T_2=T\cdot(1+d)$. Generalizing, we can derive the expected computing time of a VM when multiplexed with additional $i-1$ VMs as $T_i=T\cdot(1+d)^{i-1}$.} at BS $n$ to specify the computation rate reduction of a VM when multiplexed with another VM. With one-to-one correspondence between each VM and each user as mentioned earlier, the number of VMs hosted at a BS is equal to the number of the associated users $\sum_{k\in\mathcal{K}}x_{k,n}$. Therefore, given $\sum_{k\in\mathcal{K}}x_{k,n}$ associated users at BS $n$, the actual computation rate for execution of user $k$'s task is
\begin{align}\label{eqn:2}
F_{k,n}=f_{k,n} (1+d_n)^{1-\sum_{k\in\mathcal{K}}x_{k,n}}.
\end{align}
Here \eqref{eqn:2} indicates that the computation rate of each user decreases as the number of co-located users at a BS increases. This implies a tradeoff that accommodating more users (or consolidating more VMs) at a BS can increase multiplexing gain in parallel computing but degrades the computation rates for individual users due to the I/O interference.On the other hand, considering the finite computation capacity of a BS, we assume that the number of multiplexed VMs (or equivalently, the number of users) at a BS is bounded by a number $M_n$, i.e., $\sum_{k\in\mathcal{K}}x_{k,n}\leq M_n$.

After characterizing user's communication rate in \eqref{eqn:1} and computation rate in \eqref{eqn:2}, \mbox{we use the} offloading rate as a metric to measure the computation offloading performance. Here, the offloading rate is defined as the number of user's offloadable bits per unit time, which is \mbox{given by}
\begin{align}\label{eqn:Vkn}
R_{k,n}=\left.1\middle/\left(\frac{1}{r_{k,n}}+\frac{1}{F_{k,n}}\right)\right.,
\end{align}
where the first and second terms in the denominator are the inverse of the transmission rate  and computation rate, respectively, denoting the corresponding required time for transmitting and computing $1$-bit.

\vspace{-0.15in}
\subsection{Problem Formulation}
We consider a problem of service migration among BSs under the consideration of joint computation-and-radio resource management. Specifically, given the initial offloading-service placement, we
aim to find the optimal JMH decisions that maximize the weighted sum offloading rate while reduce the total incurred JMH cost at the same time. The problem is formulated as
\begin{alignat}{2}
(\text{P1}):\quad  \max_{\mathbf{X}} \quad &\sum_{k\in\mathcal{K}}\omega_k\sum_{n\in\mathcal{N}} x_{k,n}R_{k,n}- \lambda\sum_{k\in\mathcal{K}}\sum_{n\in\mathcal{N}}\sum_{j\in\mathcal{N}}x_{k,j}^{0} x_{k,n}c_{k,j,n} \label{eqn:ob1}
\\[0mm]
{\rm{s.t.}} \quad&\sum_{n\in\mathcal{N}}x_{k,n}=1, \quad \quad ~\forall k\in\mathcal{K}, \label{eqn:st1}\\[0mm]
&\sum_{k\in\mathcal{K}}x_{k,n}\leq M_n, \quad ~ ~\forall n\in\mathcal{N},\label{eqn:st3}\\[0mm]
&x_{k,n}\in\{0,1\}, ~\quad\quad\forall  k\in\mathcal{K},  ~\forall n\in\mathcal{N}, \label{eqn:st2}
\end{alignat}
where $\mathbf{X}=\{x_{k,n}\}$ and $\omega_k\geq 0$ denotes a weight assigned to user $k$'s offloading rate. $\lambda\geq 0$ is a weight factor for adjusting the sum offloading rate and JMH cost, which is determined by the system operator according to the system preference\footnote{To set $\lambda$, we can first measure the statistical performance of sum offloading rate and the total JMH cost for different values of $\lambda$, which is obtained by running the proposed Algorithm \ref{alg:2} for given a set of $\lambda$'s and channel realizations. Note that this step can be performed a priori. After that, we can choose a proper $\lambda$ according to the system preference.}. The objective \eqref{eqn:ob1} is to optimize the tradeoff between the weighted sum of users' offloading rates and the required JMH cost, which can be regraded as the utility by JMH. Constraint \eqref{eqn:st1} states that each user is associated with only one BS. Constraint \eqref{eqn:st3} ensures that the number of users (or VMs) served by a BS does not exceed the maximum number. Clearly, $\sum_{n\in\mathcal{N}}M_n \geq K $ should be satisfied for problem feasibility.

Due to the binary variables $\mathbf{X}$, Problem (P1) is an \emph{integer nonlinear programming} problem that is difficult to solve exactly. The brute force algorithm has a complexity of $\mathcal{O}(N^K)$, which is prohibitive when the size of the cellular network is large. To this end, we design a low-complexity and suboptimal algorithm in the next section, which is shown to have close-to-optimal performance in the simulations.

\vspace{-0.1in}
\section{Algorithm Development}
In this section, we develop an efficient algorithm to solve Problem (P1), which proceeds in two stages: First,
by integer relaxation and some mathematic manipulation, we transform Problem (P1) into a sequence of convex problems that can be optimally solved. In the second stage, as the obtained results may be fractional, we propose a new method to recover a feasible integer solution. Last, we extend the algorithm framework to the radio-resource  allocation case.

\vspace{-0.15in}
\subsection{Continuous Relaxation and Fractional Programming Transform}\label{sec:stage1}
To make Problem (P1) more tractable, we first relax the binary variable $x_{k,n}$ into $[0,1]$. Moreover, we introduce a new variable $y_n$ to replace the term $\sum_{k\in\mathcal{K}}x_{k,n}$ in \eqref{eqn:Vkn} and in \eqref{eqn:st3}, which implies the computation load (or the number of VM) that BS $n$ accommodates. Then, the relaxed Problem (P1) can be written as the following equivalent form:
\begin{align}
(\text{P1}^\prime):\quad \max_{\mathbf{X}, \mathbf{y}} \quad &\sum_{k\in\mathcal{K}}\sum_{n\in\mathcal{N}} \frac{x_{k,n}\omega_k }{\frac{1}{r_{k,n}}+\frac{(1+d_n)^{y_n-1}}{f_{k,n}}}+\sum_{k\in\mathcal{K}}\sum_{n\in\mathcal{N}}x_{k,n}z_{k,n} \label{eqn:ob2}\\
{\rm{s.t.}} \quad&\sum_{n\in\mathcal{N}}x_{k,n}=1, \quad \quad ~\forall k\in\mathcal{K}, \label{eqn:st21}\\
&\sum_{k\in\mathcal{K}}x_{k,n}\leq y_n, \quad \quad \forall n \in\mathcal{N}, \label{eqn:st22}\\
&0\leq x_{k,n}\leq1, ~\, \quad \quad \forall k\in\mathcal{K},~ \forall n \in\mathcal{N}, \label{eqn:st23}\\
&0 \leq y_n \leq M_n,\,\quad \quad \forall n \in\mathcal{N}, \label{eqn:st24}
\end{align}
where $\mathbf{y}=\{y_n\}$, and $z_{k,n}\triangleq  Z-\lambda \sum_{j\in\mathcal{N}}x_{k,j}^0c_{k,j,n}\geq 0$ in \eqref{eqn:ob2} is an aggregated term related to the cost, in which  $Z$ is a sufficiently large constant for ensuring all the $z_{k,n}$'s being non-negative, e.g., set $Z\geq  \max_{k,n}\{\lambda \sum_{j\in\mathcal{N}}x_{k,j}^0c_{k,j,n}\}$. Adding a common $Z$ to each term serves for rewriting the objective function \eqref{eqn:ob1} as a form of sum of non-negative functions, so as to meet the requirement of the sum-of-ratios algorithm design. Evidently, since $R_{k,n}$ monotonically decreases with $y_n$, the auxiliary variable $y_n$ always achieves its lower bound $\sum_{k\in\mathcal{K}}x_{k,n}$ in \eqref{eqn:st22} for optimality, i.e., the equality holds in \eqref{eqn:st22}. Due to the integer relaxation at constraints \eqref{eqn:st23}, Problem (P1$^\prime$) yields the upper bound of the original Problem (P1).

Based on the above transformation, Problem (P1$^\prime$) becomes a continuous optimization problem with the sum-of-ratios objective. According to  \cite{sum_of_ratios}, we can transform  Problem (P1$^\prime$) into an equivalent parameterized subtractive-form problem via the following theorem.
\begin{theorem}\label{theorem1}
If $(\mathbf{X}^*, \mathbf{y}^*)$ is the optimal solution to Problem (P1$^\prime$), then there exist $\bm{\alpha}^*=\{\alpha_{k,n}^*\}$, $\bm{\beta}^*=\{\beta_{k,n}^*\}$, and $\bm{\gamma}^*=\{\gamma_{k,n}^*\}$ such that $(\mathbf{X}^*, \mathbf{y}^*)$ is the optimal solution to the following parameterized problem with $(\bm{\alpha}, \bm{\beta}, \bm{\gamma})=(\bm{\alpha}^*, \bm{\beta}^*, \bm{\gamma}^*)$:
\begin{align}
(\text{P2}):\quad \max_{(\mathbf{X}, \mathbf{y)\in \mathcal{F}}} \quad &\sum_{k\in\mathcal{K}}\sum_{n\in\mathcal{N}}\alpha_{k,n}\Big[x_{k,n}\omega_k-\beta_{k,n}\Big(\tfrac{1}{r_{k,n}}+\tfrac{(1+d_n)^{y_n-1}}{f_{k,n}}\Big)\Big] \nonumber \\
&+\sum_{k\in\mathcal{K}}\sum_{n\in\mathcal{N}}(x_{k,n}z_{k,n}-\gamma_{k,n}), \label{eqn:ob3}
\end{align}
\end{theorem}
where $\mathcal{F}$ denotes the feasible solution set satisfying the constraints \eqref{eqn:st21}-\eqref{eqn:st24}. Moreover,  $(\mathbf{X}^*, \mathbf{y}^*)$ also satisfies the following conditions when $(\bm{\alpha}, \bm{\beta}, \bm{\gamma})=(\bm{\alpha}^*, \bm{\beta}^*, \bm{\gamma}^*)$,  for all $k$ and $n$:
\begin{align}
\alpha_{k,n}\Big(\tfrac{1}{r_{k,n}}+\tfrac{(1+d_n)^{y_n-1}}{f_{k,n}}\Big)-1&=0,  \label{eqn:condi1}\\
\beta_{k,n}\Big(\tfrac{1}{r_{k,n}}+\tfrac{(1+d_n)^{y_n-1}}{f_{k,n}}\Big)-x_{k,n}\omega_k&=0, \label{eqn:condi2}\\
\gamma_{k,n}-x_{k,n}z_{k,n}&=0.\label{eqn:condi3}
\end{align}
\begin{IEEEproof}
See Appendix A.
\end{IEEEproof}

Theorem \ref{theorem1} reveals that the sum-of-ratios maximization Problem (P1$^\prime$) shares the same optimal solution with the parameterized subtractive-form Problem (P2) when $(\bm{\alpha}, \bm{\beta}, \bm{\gamma})=(\bm{\alpha}^*, \bm{\beta}^*, \bm{\gamma}^*)$. Here, $(\bm{\alpha}^*, \bm{\beta}^*, \bm{\gamma}^*)$ denotes the optimal tuple of parameters that meets the system equations \eqref{eqn:condi1}-\eqref{eqn:condi3} together with its corresponding solution $(\mathbf{X}, \mathbf{y})$ to Problem (P2). Based on Theorem \ref{theorem1}, we can solve Problem (P1$^\prime$) by a two-layer iterative approach: In the inner layer, we solve the subtractive-form Problem (P2) with given $(\bm{\alpha}, \bm{\beta}, \bm{\gamma})$. Then, in the outer layer, we find the optimal $(\bm{\alpha}^*, \bm{\beta}^*, \bm{\gamma}^*)$ satisfying \eqref{eqn:condi1}-\eqref{eqn:condi3}.

\vspace{-0.15in}
\subsection{Solving Problem (P2) Given $(\bm{\alpha}, \bm{\beta}, \bm{\gamma})$}\label{sec:stage2}
In this subsection, we solve Problem (P2) for given parameters $(\bm{\alpha}, \bm{\beta}, \bm{\gamma})\succeq0$, which can be further re-expressed as
\begin{align}
\max_{(\mathbf{X}, \mathbf{y)\in \mathcal{F}}} \quad \sum_{k\in\mathcal{K}}\sum_{n\in\mathcal{N}}x_{k,n}&(\alpha_{k,n}\omega_k+z_{k,n})-\sum_{n\in\mathcal{N}}\Big[\Big(\sum_{k\in\mathcal{K}}\tfrac{\alpha_{k,n}\beta_{k,n}}{f_{k,n}}\Big)(1+d_n)^{y_n-1}\Big]\label{eqn:ob4}
\end{align}
where the objective function \eqref{eqn:ob4} is derived from \eqref{eqn:ob3} by omitting the constant terms $\sum_{k}\!\sum_{n}\!\gamma_{k,n}$ and $\sum_{k}\!\sum_{n}\!\frac{\alpha_{k,n}\beta_{k,n}}{r_{k,n}}$.

It can be readily proved that Problem \eqref{eqn:ob4} is convex because the objective function is concave and the constraints are linear. Then, the convex optimization methods can be used to solve this problem optimally. By introducing a set of Lagrangian multipliers $\bm{\mu}=\{\mu_n\}\succeq0$ associated with the constraints \eqref{eqn:st22}, the dual problem of Problem \eqref{eqn:ob4} can be expressed as
\begin{align}\label{eqn:dual}
\min_{\bm{\mu}\succeq0} \quad \theta(\bm{\mu})=\sum_{k\in\mathcal{K}} \phi_k(\bm{\mu})+\sum_{n\in\mathcal{N}} \xi_n(\mu_n),
\end{align}
where
\begin{gather}
\phi_{k} (\bm{\mu})=\left\{
\begin{aligned}
\max\limits_{\{x_{k,n}\}_{n\in\mathcal{N}}}\quad  & \sum\limits_{n\in\mathcal{N}}x_{k,n}(\alpha_{k,n}\omega_k+z_{k,n}-\mu_{n})\\
\quad{\rm{s.t.}}~~\quad &\sum\limits_{n\in\mathcal{N}}x_{k,n}=1,\\
&0\leq x_{k,n}\leq 1, \quad \forall n\in\mathcal{N},
\end{aligned}\right. \label{eqn:dualx}\\
\begin{aligned}
\xi_n(\mu_n)&=\max_{0 \leq y_n \leq M_n} ~~\mu_ny_n-\Big(\sum_{k\in\mathcal{K}}\tfrac{\alpha_{k,n}\beta_{k,n}}{f_{k,n}}\Big)(1+d_n)^{y_n-1}.
\end{aligned}\label{eqn:dualy}
\end{gather}
Since all the constraints in the convex Problem \eqref{eqn:ob4} are linear, the Slater's condition is satisfied and the strong duality holds \cite{boyd2004convex}. The primal Problem \eqref{eqn:ob4} can therefore be equivalently solved by the dual Problem \eqref{eqn:dual}.

\subsubsection{Optimal JMH Policy in Dual Domain}
We can observe that the dual function $\theta(\bm{\mu})$ has a decomposable structure. Specifically, given $\bm{\mu}$, $\theta(\bm{\mu})$ can be determined by solving $K$ independent subproblems \eqref{eqn:dualx}, where each user $k$ individually makes its own JMH decision $\{x_{k,n}\}_{n\in\mathcal{N}}$ over the BSs, and at the same time $N$ independent subproblems \eqref{eqn:dualy}, where each BS $n$ optimizes its own computational load $y_n$.

To solve the JMH subproblem \eqref{eqn:dualx} for each user, we have the following observation:
\begin{remark}[JMH Revenue]\label{Remark1}
The value of $(\alpha_{k,n}\omega_k+z_{k,n}-\mu_n)$ in subproblem \eqref{eqn:dualx} can be interpreted as the revenue of user $k$ when its service is migrated to BS $n$.
Specifically, with $\alpha_{k,n}$ referred as to the  offloading rate in each iteration [see \eqref{eqn:condi1}], $\left(\alpha_{k,n}\omega_k+z_{k,n}\right)$ represents the profit obtained from BS $n$, consisting of the weighted offloading rate $\omega_k\alpha_{k,n}$ and the modified cost $z_{k,n}$. On the other hand, the Lagrangian multiplier $\mu_n$ is the price of BS $n$ to provide service. Therefore, the difference between the profit and the payment, $\left(\omega_k\alpha_{k,n}+z_{k,n}-\mu_n\right)$, can be measured as the revenue of user $k$ obtained from BS $n$.
\end{remark}

Based on Remark \ref{Remark1}, the objective of subproblem \eqref{eqn:dualx} can be interpreted as maximizing the revenue of user $k$ over all the BSs. Through a direct observation, each subproblem \eqref{eqn:dualx} always has an optimal \textit{binary} solution $\{x_{k,n}^*\}$:
\begin{align}\label{eqn:xkn_op}
x_{k,n}^*=
\begin{cases}
1, &\text{if $n=n_k^{*}=\argmax\limits_{n^\prime\in\mathcal{N}} \left\{ \alpha_{k,n^\prime}\omega_k+z_{k,n^\prime}-\mu_{n^\prime}\right\}$},\\
0, & \text{otherwise},
\end{cases}
\end{align}
i.e., each user selects one BS with the highest revenue. Note that when there are multiple maximizers $n_k^*$'s, user can choose any one of them without affecting the value of dual function.

For subproblem \eqref{eqn:dualy}, the optimal amount of load $y_n^*$ at BS $n$ can be obtained via differentiating $\xi_n(\mu_n)$ in \eqref{eqn:dualy} with respect to $y_n$ and letting the result be zero:
\begin{align}\label{eqn:yn_op}
y_{n}^*=
\begin{cases}
\min\left\{\frac{\ln \mu_n-\ln q_n}{\ln(1+d_n)}+1,M_n\right\} &~~\text{if~}\mu_n \geq \mu_n^{\text{min}}\triangleq\frac{q_n}{1+d_n},\\
0, &~~\text{if~} 0\leq \mu < \mu_n^{\text{min}},
\end{cases}
\end{align}
where $q_n = (\sum_{k\in\mathcal{K}}\frac{\alpha_{k,n}\beta_{k,n}}{f_{k,n}})\ln(1+d_n)$.



For the dual Problem \eqref{eqn:dual}, we use the subgradient method to find the optimal Lagrangian multipliers $\bm{\mu}^*$, in which each $\mu_n$ is updated as
\begin{align}\label{eqn:subgradi}
\mu_n^{t+1}=\Big[\mu_n^{t}-\epsilon^{t} \Big(y_n^{t}-\sum_{k\in\mathcal{K}}x_{k,n}^{t}\Big)\Big]^+,
\end{align}
for $t=1, 2, ...$, where $[\cdot]^+\!\triangleq\!\max\{\cdot,0\}$ and $\epsilon^{t}$ is the step size chosen in iteration $t$. In this paper, we adopt a harmonic series step size $\epsilon^{t}=\epsilon/(t+1)$, $t=1, 2, \cdots$. $\epsilon>0$ is a properly designed constant. Since the primal problem is a convex problem satisfying Slater's condition, the subgradient method in \eqref{eqn:subgradi} operated with the above step-size rule guarantees the convergence to an optimal dual solution $\bm{\mu}^*$ and the primal optimal value \cite{two_known}.

\subsubsection{Optimal Primal Solution Recovery for Problem (P2)}\label{sssec:step2}
Although the optimal $\bm{\mu}^*$ is obtained by the above subgradient method, its associated solution $(\mathbf{X}(\bm{\mu}^*), \mathbf{y}(\bm{\mu}^*))$ by \eqref{eqn:xkn_op} and \eqref{eqn:yn_op} may not be optimal and can even be infeasible for the primal Problem (P2). This is because the dual subgradient method does not guarantee to find an optimal primal solution even for the convex problem satisfying strong duality, unless the dual function $\theta(\bm{\mu})$ is differentiable at $\bm{\mu}^*$ \cite{two_known,primal_converge}. In our Problem (P2), it arises from the fact that the dual subproblem \eqref{eqn:dualx} is a linear programming (LP) problem. When there exists a user that has more than one BS with the same maximum revenue at $\bm{\mu}^*$, the binary-form solution in \eqref{eqn:xkn_op} is not a unique solution to the dual Problem \eqref{eqn:dual} such that it may not be primal optimal to Problem (P2)
(see \cite[Proposition 7]{primal_converge}). However, the optimal solution to the inner-layer Problem (P2) with given $(\bm{\alpha}, \bm{\beta}, \bm{\gamma})$ is required for ensuring the convergence of sum-of-ratios algorithm \cite{sum_of_ratios}.

To address this issue, we adopt the average procedure \cite{primal_converge} to recover the primal solution. The idea behind is to reconstruct an approximate primal feasible solution by a weighted convex combination of the previous primal iterates $(\{x_{k,n}^t\}, \{y_n^t\})$ obtained by \eqref{eqn:xkn_op} and \eqref{eqn:yn_op}, which is shown to converge an optimal primal solution.
\begin{theorem}[Primal Convergence]\label{theorem2}
Consider the primal-and-dual iteration scheme [\eqref{eqn:xkn_op}, \eqref{eqn:yn_op}, and \eqref{eqn:subgradi}] for  Problem (P2) and that  we recursively average the primal iterates $(\{x_{k,n}^t\}, \{y_n^t\})$ obtained by \eqref{eqn:xkn_op} and \eqref{eqn:yn_op} as follows:
\begin{align}
\bar{x}_{k,n}^{t}&=\left(1-\frac{t^\nu}{\sum_{s=1}^{t}s^\nu}\right)\; \bar{x}_{k,n}^{t-1}+\frac{t^\nu}{\sum_{s=1}^{t}s^\nu}\; x_{k,n}^{t},  \label{eqn:primalrecover1}\\ \bar{y}_{n}^{t}&=\left(1-\frac{t^\nu}{\sum_{s=1}^{t}s^\nu}\right)\; \bar{y}_{n}^{t-1}+\frac{t^\nu}{\sum_{s=1}^{t}s^\nu}\; y_{n}^{t},  \label{eqn:primalrecover2}
\end{align}
for $t=1, 2, ...$, where $\nu\!>\!0$ is a proper constant for controlling weights. Then, $\bar{x}_{k,n}^{t}\rightarrow x_{k,n}^*$, $\forall k,n$, and $\bar{y}_n^t\rightarrow y_n^*$, $\forall n$, i.e., converge to the optimal solution of Problem (P2).
\end{theorem}
\begin{IEEEproof}
See \cite[Theorem 2]{primal_converge}.
\end{IEEEproof}
We summarize the detailed procedures of solving the inner-layer Problem (P2) in Algorithm \ref{alg:1}.

\begin{algorithm}[t]
\caption{Optimal Algorithm for Solving Problem (P2)}
\begin{algorithmic}[1]\label{alg:1}
\REQUIRE $(\bm{\alpha}, \bm{\beta}, \bm{\gamma})$.
\STATE Initialize $\{\mu_n\geq 0\}$.
\REPEAT
\STATE Compute $\{x_{k,n}\}$ and $\{y_n\}$ for given $\bm{\mu}$ according to \eqref{eqn:xkn_op} and \eqref{eqn:yn_op}, respectively.
\STATE Update $\bm{\mu}$ based on subgradient method in \eqref{eqn:subgradi}.
\STATE Update primal variables $\{\bar{x}_{k,n}\}$ and $\{\bar{y}_n\}$ according to \eqref{eqn:primalrecover1} and \eqref{eqn:primalrecover2}, respectively.
\UNTIL $\bm{\mu}$ converges.
\RETURN $x_{k,n}^*=\bar{x}_{k,n}$, $\forall k, n$ and $y_n^*=\bar{y}_n$, $\forall n$.
\ENSURE $(\mathbf{X}^*$, $\mathbf{y}^*)$ for given $(\bm{\alpha}, \bm{\beta}, \bm{\gamma})$.
\end{algorithmic}
\end{algorithm}

\vspace{-0.15in}
\subsection{Finding Optimal Parameters $(\bm{\alpha}^*, \bm{\beta}^*, \bm{\gamma}^*)$}\label{sec:stage3}
After obtaining the optimal $(\mathbf{X}^*, \mathbf{y}^*)$ for given $(\bm{\alpha}, \bm{\beta}, \bm{\gamma})$ in above subsection,  we develop an algorithm to find the optimal $(\bm{\alpha}^*, \bm{\beta}^*, \bm{\gamma}^*)$ for solving Problem ($\text{P1}^\prime$). For notational brevity, we denote $q_{k,n}\triangleq \tfrac{1}{r_{k,n}}+\tfrac{(1+d_n)^{y_n-1}}{f_{k,n}}$ and define some functions (for all $k$ and $n$) as follows:
\begin{align}
\psi_{k,n}^{1}(\alpha_{k,n})&=\alpha_{k,n}q_{k,n}-1,\\
\psi_{k,n}^{2}(\beta_{k,n})&=\beta_{k,n}q_{k,n}-x_{k,n}\omega_k,\\
\psi_{k,n}^{3}(\gamma_{k,n})&=\gamma_{k,n}-x_{k,n}z_{k,n},
\end{align}
where $(\{x_{k,n}\}, \{y_n\})$ is the inner-layer optimal solution obtained by Algorithm \ref{alg:1}.

According to \cite[Theorem 3.1]{sum_of_ratios}, the unique optimal solution of $(\bm{\alpha}^*, \bm{\beta}^*, \bm{\gamma}^*)$ is achieved if and only if $\psi_{k,n}^{i}=0$, $\forall k,n$ and $\forall i\in\{1,2,3\}$, as in conditions \eqref{eqn:condi1}-\eqref{eqn:condi3}. We employ the modified Newton method \cite{sum_of_ratios} to update $\alpha_{k,n}, \beta_{k,n}$ and $\gamma_{k,n}$ to meet above conditions. Specifically, the parameters (for all $k$ and $n$) are point-wisely updated as
\begin{align}
\alpha_{k,n}^{l+1}&=\left(1-\zeta^{l}\right)\alpha_{k,n}^{l}+ \zeta^{l}\frac{1}{q_{k,n}},\label{eqn:parameterupdate1} \\
\beta_{k,n}^{l+1}&=\left(1-\zeta^{l}\right)\beta_{k,n}^{l}+ \zeta^{l}\frac{x_{k,n}\omega_k}{q_{k,n}}, \label{eqn:parameterupdate2} \\
\gamma_{k,n}^{l+1}&=\left(1-\zeta^{l}\right)\gamma_{k,n}^{l}+ \zeta^{l} x_{k,n}z_{k,n}, \label{eqn:parameterupdate3}
\end{align}
where $l$ is the iteration index for the sum-of-ratios algorithm. $\zeta^{l}$ is the step size at  iteration $l$ chosen via the following manner. Let $m_l$ be the smallest integer among $m\in\{0,1,...\}$ satisfying
\begin{small}
\begin{align}\label{eqn:mk}
&\sum_{k\in\mathcal{K}}\!\sum_{n\in\mathcal{N}}\!\bigg\{\!\Big|\psi_{k,n}^{1}\Big(\!(1\!-\!\rho^{m})\alpha_{k,n}^{l}\!+\! \rho^{m}\frac{1}{q_{k,n}}\!\Big)\!\Big|^2\!+\!\Big|\psi_{k,n}^{2}\Big(\!(1\!-\!\rho^{m})\beta_{k,n}^{l}\!+\! \rho^{m}\frac{x_{k,n}\omega_k}{q_{k,n}}\!\Big)\Big|^2 \bigg. \nonumber \\
&\bigg.\!+\!\Big|\psi_{k,n}^{3}\Big(\!(1\!-\!\rho^{m})\gamma_{k,n}^{l}\!+\! \rho^{m} x_{k,n}z_{k,n}\!\Big)\!\Big|^2\!\bigg\}\leq \!(1\!-\!\varepsilon\rho^{m})\!\sum_{k\in\mathcal{K}}\!\sum_{n\in\mathcal{N}}\!\Big(|\psi_{k,n}^{1}(\alpha_{k,n}^{l})|^2\!+\!|\psi_{k,n}^{2}(\beta_{k,n}^{l})|^2\!+\!|\psi_{k,n}^{3}(\gamma_{k,n}^{l})|^2\!\Big),
\end{align}
\end{small}
where $\varepsilon\in(0,1)$ and $\rho\in(0,1)$. We set $\zeta^{l}=\rho^{m_l}$ at the $l$-th iteration.

As indicated in \cite{sum_of_ratios}, the sum-of-ratios iterative algorithm can converge to the global optimum of Problem (P1$^\prime$) if the inner-layer Problem (P2) for given $(\bm{\alpha}, \bm{\beta}, \bm{\gamma})$ is optimally solved and the outer-layer update of $(\bm{\alpha}, \bm{\beta}, \bm{\gamma})$ is via the modified Newton method \eqref{eqn:parameterupdate1}-\eqref{eqn:parameterupdate3}. Evidently, the global optimum of Problem (P2) can be guaranteed by Algorithm 1 due to its convexity. Thus, our proposed sum-of-ratios algorithm can achieve the global optimal solution to Problem (P1$^\prime$).


\vspace{-0.1in}
\subsection{Integer Recovery for Problem (P1)}\label{sec:stage4}
Let ($\mathbf{X}^\prime\!, \mathbf{y}^\prime$) denote the optimal solution to Problem ($\text{P1}^\prime$). As explained in Section \ref{sec:stage2},  ($\mathbf{X}^\prime, \mathbf{y}^\prime$) may be fractional due to the possibility that the binary-form solution in \eqref{eqn:xkn_op} is not the primal optimal to Problem (P2). Therefore, in this subsection, we discuss the integer recovery on JMH decisions $\mathbf{X}^\prime$ to finalize solving Problem (P1). There are two major challenges in recovery for our problem instance. First of all, the recovery operation needs to guarantee the obtained result still meeting the hard constraints \eqref{eqn:st1} and \eqref{eqn:st2} of Problem (P1). Second, since the user's offloading rate $R_{k,n}$ is a function of the sum of users' decisions $\sum_{k}x_{k,n}$ [see in \eqref{eqn:Vkn}], rounding $x_{k,n}$ without considering this correlation may accumulate a significant variance in $\sum_{k}x_{k,n}$, which in turn affects $\{R_{k,n}\}$ greatly in the objective of Problem (P1) and incurs high performance loss. In order to recover a feasible decision solution with less rounding loss, we propose an effective rounding method that captures the problem structure. The key idea is to utilize an important property of Problem (P1) given any feasible integer  $\mathbf{y}$, which is described as follows.
\begin{theorem}\label{pro1}
Define $\mathcal{Y}\triangleq\{\mathbf{y}\in \mathbb{Z}^{N}|\sum_{n\in\mathcal{N}}y_n=K, \text{and~} 0\leq y_n\leq M_n, \forall n \in \mathcal{N}\}$ as the feasible integer set of $\mathbf{y}$, where $\mathbb{Z}^{N}$ denotes the integer set. For any given $\mathbf{y}\in \mathcal{Y}$, Problem (P1) is reduced into an integer linear programming (ILP) problem,  expressed as
\begin{align}
(\text{P3}):\quad \max_{\mathbf{X}} \quad &\sum_{k\in\mathcal{K}}\sum_{n\in\mathcal{N}} x_{k,n}u_{k,n}(y_n)  \label{eqn:ob5}\\
{\rm{s.t.}} \quad&\sum_{n\in\mathcal{N}}x_{k,n}=1, \quad\quad~\forall k\in\mathcal{K}, \label{eqn:st51}\\
&\sum_{k\in\mathcal{K}}x_{k,n}= y_n, \quad\quad~ \forall n\in\mathcal{N}, \label{eqn:st52} \\
&x_{k,n}\in\{0,1\}, ~ \quad\quad~ \forall k\in\mathcal{K},~\forall n\in\mathcal{N}, \label{eqn:st53}
\end{align}
where $u_{k,n}(y_n)\triangleq\frac{\omega_k}{\frac{1}{r_{k,n}}+\frac{(1+d_n)^{y_n-1}}{f_{k,n}}}-\lambda \sum_{j\in\mathcal{N}}x_{k,j}^0c_{k,j,n}$ is pre-calculated for the given $\mathbf{y}$. And \makebox{Problem (P3)} is equivalent to the linear assignment problem (LAP).
\end{theorem}
\begin{IEEEproof}
See \cite[Theorem 1]{graph_construct}.
\end{IEEEproof}

According to Theorem \ref{pro1}, we can map Problem (P1) with any given $\mathbf{y}\in\mathcal{Y}$ into an equivalent LAP problem. It is well-known that the LAP problem is a special linear integer programming problem with a nice combinatorial property that its integer-relaxed problem always has an integer optimal solution, i.e., LAP is equivalent to its continuous relaxation. Also, the famous Hungarian algorithm \cite{hungarian} can provide an optimal solution to LAP in a polynomial complexity of $\mathcal{O}(K^3)$. As a result, the optimal $\mathbf{X}$ to Problem (P1) can be efficiently obtained once $\mathbf{y}\in\mathcal{Y}$ is determined.

Next, we turn to construct an effective $\mathbf{y}\in\mathcal{Y}$ by rounding the fractional-optimal $\mathbf{y}^\prime$. Note that $\mathbf{y}^\prime$ satisfies $\sum_{n\in\mathcal{N}}y_n^\prime\!=\!K$ because of the necessarily optimal condition $\sum_{k\in\mathcal{K}}x_{k,n}^\prime\!=\!y_n^\prime$ in \eqref{eqn:st22}, and $\lceil y_n^\prime\rceil \leq M_n$, $\forall n$, since $y_n^\prime \leq M_n$ by \eqref{eqn:st24} and $M_n$ is integral, where $\lfloor \cdot \rfloor/\lceil \cdot \rceil$ denotes the floor/ceil operation. Let $s\triangleq K\!-\!\sum_{n\in \mathcal{N}}\lfloor y_n^\prime \rfloor$ and $\hat{\mathbf{y}}\triangleq \{\hat{y}_n\}$ be the recovered integer solution. We round $\mathbf{y}^\prime$ to construct $\hat{\mathbf{y}}$ by setting $\hat{y}_n = \lceil y_n^\prime \rceil$ for $s$ BSs with the maximal value of $(y_n^\prime - \lfloor y_n^\prime\rfloor)$ and setting $\hat{y}_n = \lfloor y_n^\prime \rfloor$ for the rest of BSs. Mathematically, the recovered $\hat{y}_n$ is given by
\begin{align}\label{eqn:round}
\hat{y}_n=\begin{cases}
\lceil y_n^\prime\rceil, &\text{if~}  y_n^\prime - \lfloor y_n^\prime\rfloor \text{~is one of the $s$ largest},\\
\lfloor y_n^\prime\rfloor, &\text{otherwise}.
\end{cases}
\end{align}
\begin{proposition}\label{pro10} $\hat{\mathbf{y}}$ constructed by rule \eqref{eqn:round} satisfies:
\begin{enumerate}
\item [a)]$\hat{\mathbf{y}}\in \mathcal{Y}$, i.e., $\hat{\mathbf{y}}$ is an integer vector meeting $\sum_{n\in\mathcal{N}}y_n=K$ and $0\leq y_n\leq K$, $\forall n\in\mathcal{N}$;
\item [b)] $\hat{\mathbf{y}}\in\arg\min_{\mathbf{y}\in\mathcal{Y}}\|\mathbf{y}-\mathbf{y}^\prime\|_q$, with $q\!\geq\!1$, i.e., $\hat{\mathbf{y}}$ is one of the closest integer vectors in set $\mathcal{Y}$ to the fractional-optimal $\mathbf{y}^\prime$,  for any norm $q\geq 1$.
\end{enumerate}
\end{proposition}
\begin{IEEEproof}
See Appendix B.
\end{IEEEproof}

Compared with the method that directly rounds $x_{k,n}$ and incurs a unstable rounding error on $|\hat{y}_n-y_n^\prime|$, the rounding rule \eqref{eqn:round} generates a feasible integer $\mathbf{y}\in\mathcal{Y}$ with the smallest rounding error $\|\mathbf{y}-\mathbf{y}^\prime\|_q$. Moreover, as the optimal $\mathbf{X}$ with given $\mathbf{y}\in\mathcal{Y}$ can be optimally solved by Hungarian algorithm, it can be perceived that our recovery method has lower performance loss than that of the direct rounding.

\begin{algorithm}[t]
\caption{Whole Algorithm for Solving Problem (P1)}
\begin{algorithmic}[1]\label{alg:2}
\STATE Initialize $(\bm{\alpha}, \bm{\beta}, \bm{\gamma})\succeq0$.
\REPEAT
\STATE Given $(\bm{\alpha}, \bm{\beta}, \bm{\gamma})$, obtain the optimal solution $(\mathbf{X}^\prime, \mathbf{y}^\prime)$ to Problem (P2) by Algorithm \ref{alg:1}.
\STATE Update $(\bm{\alpha}, \bm{\beta}, \bm{\gamma})$ using \eqref{eqn:parameterupdate1}, \eqref{eqn:parameterupdate2}, and \eqref{eqn:parameterupdate3}.
\UNTIL $\sum_{i=1}^3\sum_{k\in\mathcal{K}}\sum_{n\in\mathcal{N}}|\psi_{k,n}^{i}|^2<\epsilon$, where $\epsilon$ controls accuracy.
\STATE Round $\mathbf{y}^\prime\rightarrow \hat{\mathbf{y}}$ by rule \eqref{eqn:round}.
\STATE Given $\hat{\mathbf{y}}$, obtain the optimal solution $\hat{\mathbf{X}}$ by solving Problem (P3).
\ENSURE $\hat{\mathbf{X}}$.
\end{algorithmic}
\end{algorithm}

Based on the discussions above, we present the whole algorithm procedures of solving Problem (P1) in Algorithm \ref{alg:2}. Its computational complexity is dominated by the sum-of-ratios algorithm in Steps 2-5 and solving the LAP problem at the rounding stage in Step 7. The sum-of-ratios algorithm is an iterative method that repeatedly solves the parameterized Problem (P2) by Algorithm \ref{alg:1} and updates auxiliary parameters until convergence. The complexity of \makebox{Algorithm \ref{alg:1}} is $\mathcal{O}(NK/\delta^2)$, where the complexity of computing $(\mathbf{X}, \mathbf{y})$ per iteration is $\mathcal{O}(NK)$ and the subgradient method iterates $\mathcal{O}(1/\delta^2)$ to converge, given a solution accuracy of $\delta>0$ \cite{boyd2004convex}. Thus, the total complexity of sum-of-ratios algorithm is $\mathcal{O}(T_1NK/\delta^2)$, where $T_1$ is the number of sum-of-ratio iterations and is independent of the amount of variables and fractional functions \cite{sum_of_ratios}. Solving the LAP Problem (P3) using Hungarian algorithm is of complexity $O(K^3)$. Therefore, Algorithm \ref{alg:2} has the total complexity of $O(T_1NK/\delta^2+K^3)$.

\vspace{-0.1in}
\subsection{Extension: JMH with Radio Resource Allocation}\label{ssec:BA}
In the previous sections, we consider the full frequency reuse scheme in multiuser's offloading to reduce the complexity of analysis. Allocating users with different sub-bands is however necessary in large systems to mitigate their mutual interference. To this end, in this subsection we consider radio resource allocation into the JMH design.

Consider that the spectrum of BSs do not overlap each other and each BS allocates its time and frequency radio resources, which is known as physical resource blocks (RBs), to the associated users in an orthogonal manner. We denote $\eta_{k,n}$ as the spectral efficiency in uplink transmission between user $k$ and BS $n$, and $b_{k,n}$ as the amount of RBs allocated by BS $n$ to user $k$. Based on above assumptions, the achievable uplink transmission rate of user $k$ to BS $n$ is rewritten by $r_{k,n}=b_{k,n} \eta_{k,n}$ and the JMH Problem (P1) with radio resource allocation can be formulated as
\begin{align}
\max_{\mathbf{X},\mathbf{B}} \quad &\sum_{k\in\mathcal{K}}\sum_{n\in\mathcal{N}}\frac{x_{k,n}\omega_k}{\frac{1}{b_{k,n}\eta_{k,n}}+\frac{1}{F_{k,n}}}- \lambda\sum_{k\in\mathcal{K}}\sum_{n\in\mathcal{N}}\sum_{j\in\mathcal{N}}x_{k,j}^{0} x_{k,n}c_{k,j,n} \label{eqn:bwob1}
\\
{\rm{s.t.}} \quad&\sum_{k\in\mathcal{K}}x_{k,n} b_{k,n} \leq B_n,\quad\quad \quad \quad  \forall n\in\mathcal{N}, \label{eqn:bwst1}\\
&\sum_{n\in\mathcal{N}} x_{k,n}=1,\quad \quad \quad\quad \; \quad \quad \forall n\in\mathcal{K}, \\
&\sum_{k\in\mathcal{K}}x_{k,n}\leq M_n, \quad \quad \quad \quad ~~  \quad\forall n\in\mathcal{N}, \\
&x_{k,n}\in\{0,1\}, ~~b_{k,n}\geq 0, \quad~~\, \ \forall k\in\mathcal{K},~\forall n\in\mathcal{N},
\end{align}
where $\mathbf{B}=\{b_{k,n}\}$. Note that if $b_{k,n}=0$, the offloading rate $R_{k,n}$ in \eqref{eqn:bwob1} is equal to zero. \eqref{eqn:bwst1} is the radio resource capacity constraint on each BS, with $B_n$ denoting the total amount of RBs at BS $n$. It is easy to check that constraint \eqref{eqn:bwst1} can be equivalently re-written as $\sum_{k\in\mathcal{K}}b_{k,n}\leq B_n$ since $x_{k,n}=0$ in \eqref{eqn:bwob1} would enforce  $b_{k,n}=0$.

We can extend the proposed Algorithm \ref{alg:2} to solve Problem \eqref{eqn:bwob1}. Specifically, in the stage of sum-of-ratios algorithm design, the integer-relaxed Problem \eqref{eqn:bwob1} can be addressed by solving a sequence of subtractive-form problems with given auxiliary parameters $(\bm{\alpha}, \bm{\beta}, \bm{\gamma})$, which are convex problems expressed as
\begin{align}
\max_{\substack{(\mathbf{X}, \mathbf{y})\in \mathcal{F}\\ \mathbf{B}\in \mathcal{S}}} \quad R=&\sum_{k\in\mathcal{K}}\sum_{n\in\mathcal{N}}\alpha_{k,n}\bigg(x_{k,n}\omega_k-\beta_{k,n}\Big(\frac{1}{b_{k,n}\eta_{k,n}}+\frac{(1+d_n)^{y_n-1}}{f_{k,n}}\Big)\bigg) \nonumber \\
&~+\sum_{k\in\mathcal{K}}\sum_{n\in\mathcal{N}}(x_{k,n}z_{k,n}-\gamma_{k,n}), \nonumber \\
=&\sum_{k\in\mathcal{K}}\sum_{n\in\mathcal{N}}x_{k,n}(\alpha_{k,n}\omega_k+z_{k,n})-\sum_{n\in\mathcal{N}}\Big[\Big(\sum_{k\in\mathcal{K}}\frac{\alpha_{k,n}\beta_{k,n}}{f_{k,n}}\Big)(1+d_n)^{y_n-1}\Big]\nonumber \\
&~-\sum_{k\in\mathcal{K}}\sum_{k\in\mathcal{N}}\frac{\alpha_{k,n}\beta_{k,n}}{b_{k,n}\eta_{k,n}} - \sum_{k\in\mathcal{K}}\sum_{k\in\mathcal{N}}\gamma_{k,n} \label{eqn:bwob2}
\end{align}
where  $\mathcal{S}\triangleq \{\mathbf{B}|\sum_{k\in\mathcal{K}}b_{k,n}\leq B_n, b_{k,n}\geq 0\}$. $y_n$, $z_{k,n}$, and $\mathcal{F}$,  as specified in Section \ref{sec:stage1},  are respectively the auxiliary variable, the modified JMH cost, and the set of $(\mathbf{X}, \mathbf{y})$ satisfying constraints \eqref{eqn:st21}-\eqref{eqn:st24}.

It can be seen that for given $(\bm{\alpha}, \bm{\beta}, \bm{\gamma})$, Problem \eqref{eqn:bwob2} can be solved optimally by solving two separate problems. The first problem of optimizing $(\mathbf{X}, \mathbf{y})$ is identical to Problem \eqref{eqn:ob4} and can be solved by Algorithm \ref{alg:1}. For the second problem of optimizing $\mathbf{B}$, the optimal RB allocation $b_{k,n}^*$ can be easily obtained as
\begin{align}
b_{k,n}^*=B_n \frac{(\alpha_{k,n}\beta_{k,n}/\eta_{k,n})^{1/2}}{\sum_{k\in\mathcal{K}}(\alpha_{k,n}\beta_{k,n}/\eta_{k,n})^{1/2}}.
\end{align}

In outer layer of updating $(\bm{\alpha}, \bm{\beta}, \bm{\gamma})$, we can use the same modified Newton method in \eqref{eqn:parameterupdate1}-\eqref{eqn:parameterupdate3} to find the optimal $(\bm{\alpha}^*, \bm{\beta}^*, \bm{\gamma}^*)$. Therefore, the sum-of-ratios algorithm design for solving resource allocation included Problem \eqref{eqn:bwob1} is almost the same as the original one except the extra computation of $b_{k,n}^*$ in each iteration.

In the stage of integer recovery of $\mathbf{X}$ as well as finding its corresponding optimal $\mathbf{B}^*$, we first apply the rounding rule \eqref{eqn:round} to recover an $\mathbf{y}\in \mathcal{Y}$. Then, given the recovered $\mathbf{y}$, the residual Problem \eqref{eqn:bwob1} is expressed as
\begin{align}
\quad\quad\quad\quad\max_{\mathbf{X},\mathbf{B}} \quad &\sum_{k\in\mathcal{K}}\sum_{n\in\mathcal{N}}x_{k,n} V_{k,n}(b_{k,n}) \label{eqn:bwob3}\\
{\rm{s.t.}} \quad&\sum_{k\in\mathcal{K}}x_{k,n} b_{k,n} \leq B_n,\quad\quad  \forall n\in\mathcal{N},  \label{eqn:bwob3st1} \\
&\sum_{n\in\mathcal{N}} x_{k,n}=y_n,\quad \quad\quad ~~ \forall k\in\mathcal{K}, \label{eqn:bwob3st2}\\
&\sum_{k\in\mathcal{K}} x_{k,n}=1\quad \quad\quad\quad~~\forall n\in\mathcal{N}, \label{eqn:bwob3st3}\\
&x_{k,n}\in\{0,1\}, ~~b_{k,n}\geq 0, \quad \forall k\in\mathcal{K},~\forall n\in\mathcal{N}, \label{eqn:bwob3st4}
\end{align}
where $V_{k,n}(b_{k,n})\triangleq \frac{\omega_k}{\frac{1}{b_{k,n}\eta_{k,n}}+\frac{(1+d_n)^{y_n-1}}{f_{k,n}}}-\lambda \sum_{j\in\mathcal{N}}x_{k,j}^0c_{k,j,n}$ is a concave function of $b_{k,n}$. Compared with Problem (P3), Problem \eqref{eqn:bwob3st1} is coupled with RB allocation and more challenging to solve. Fortunately,  we can leverage the analysis of Problem (P3) and the Lagrangian relaxation method to offer an effective solution for Problem \eqref{eqn:bwob3st1}. Specifically, let $\bm{\nu}=\{\nu_n\}\succeq0$ be the Lagrangian multipliers associated with constraint \eqref{eqn:bwob3st1}. For given $\bm{\nu}$, we consider the relaxed problem
\begin{align}
Z(\bm{\nu})~\triangleq~\max_{\mathbf{X},\mathbf{B}} ~&\sum_{k\in\mathcal{K}}\sum_{n\in\mathcal{N}}x_{k,n}\left[ V_{k,n}(b_{k,n})-\nu_n b_{k,n}\right]+\sum_{n\in\mathcal{N}}\nu_n B_n, \label{eqn:bwob4} \quad  {\rm{s.t.}}~ \eqref{eqn:bwob3st2}-\eqref{eqn:bwob3st4}.
\end{align}
The optimal $b_{k,n}^*(\nu_n)$ in Problem \eqref{eqn:bwob4} can be determined by
\begin{align}
b_{k,n}^*(\nu_n)\triangleq \argmax_{b_{k,n}\geq0}\{V_{k,n}(b_{k,n})-\nu_n b_{k,n}\} = \frac{f_{k,n}}{(1+d_n)^{y_n-1}}\left[ \sqrt{\frac{\omega_k}{\nu_n}}-\frac{1}{\eta_{k,n}}\right]^+.
\end{align}
Let $U_{k,n}(\nu_{n})\triangleq V_{k,n}(b_{k,n}^*(\nu_n))-\nu_n b_{k,n}^*(\nu_n)$. By plugging $b_{k,n}^*(\nu_n)$ into \eqref{eqn:bwob4}, we have
\begin{align}
\max_{\mathbf{X}} \quad &\sum_{k\in\mathcal{K}}\sum_{n\in\mathcal{N}}x_{k,n} U_{k,n}(\nu_n) +\sum_{n\in\mathcal{N}}\nu_n B_n, \label{eqn:bwob5} \\
{\rm{s.t.}}\quad & \eqref{eqn:bwob3st2}-\eqref{eqn:bwob3st3},~~ x_{k,n}\in\{0,1\}, \quad\forall k\in\mathcal{K}, ~\forall n\in\mathcal{N},  \nonumber
\end{align}
which is a LAP problem like Problem (P3) (see Theorem \ref{pro1}) and similarly can be solved by Hungarian algorithm. The optimal Lagrangian multiplier $\bm{\nu}^*$ to the dual problem $\min_{\bm{\nu}\succeq0} Z(\bm{\nu})$ can be found using the subgradient method. Note that due to the non-convexity of Problem \eqref{eqn:bwob3}, the optimal $(\mathbf{X}^*, \mathbf{B}^*)$ obtained in dual domain may not be the primal optimum, meaning that the duality gap exists. However, the proposed dual-based algorithm is of low complexity and yields to a good solution to the primal Problem \eqref{eqn:bwob3} in some sense.

The complexity of the modified Algorithm \ref{alg:2} for solving Problem \eqref{eqn:bwob1} is $\mathcal{O}(T_1NK/\delta^2+(NK+K^3)/\delta_2^2)$, where $\mathcal{O}(NK+K^3)/\delta_2^2)$ is the complexity of the lagrangian relaxation method, including $\mathcal{O}(NK)$ and $\mathcal{O}(K^3)$ for determining $\mathbf{B}$ and $\mathbf{X}$ in each iteration and $\mathcal{O}(\frac{1}{\delta_2^2})$ for subgradient method convergence.

\vspace{-0.1in}
\section{Hotspot Mitigation Case}

\begin{figure}[t]
\setlength{\abovecaptionskip}{0.4cm}
\setlength{\belowcaptionskip}{-0.4cm}
\begin{centering}
\includegraphics[width=0.8\linewidth]{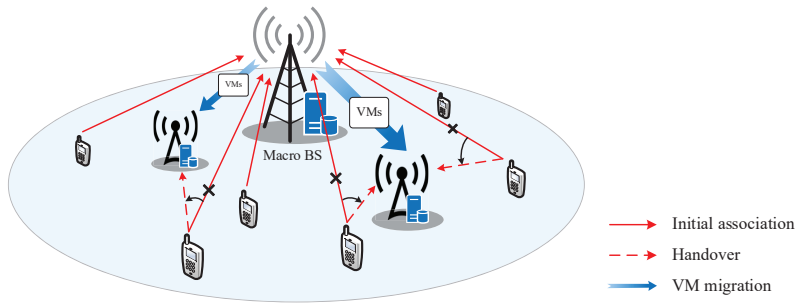}
\vspace{-0.1in}
\caption{A hotspot mitigation scenario, where an overloaded macro-BS migrates some users' services to small-BSs by JMH. }\label{fig:2}
\end{centering}
\end{figure}
In this section, we consider the JMH design for a hotspot mitigation scenario as depicted in Fig \ref{fig:2}, where a macro-BS distributes its load among $N$ idle small-BSs in a small cell. Specifically, the macro-BS, denoted by BS $0$, initially hosts all the $K$ users' services and attempts to migrate some of them to $N$ idle small-BSs for alleviating its load. Let $\mathcal{N}^+=\mathcal{N}\cup \{0\}$ denote the set of all the BSs. To facilitate exposition, we assume the users associated with the same BS $n\in\mathcal{N}^+$ have the average transmission rates and computation rates, i.e., $r_{k,n}=r_n$ and $f_{k,n}=f_{n}$, $\forall k\in \mathcal{K}$. Also, the JMH cost from BS $0$ to BS $n\in\mathcal{N}$ is assumed to be identical for each user, i.e., $c_{k,0,n}=c_{n}$, $\forall k\in \mathcal{K}$. Under the assumptions, Problem (P1) reduces to a problem of determining the number of services (or users) allocated to each BS $n\in\mathcal{N}^+$, which can be formulated as
\begin{align}
(\text{P4}):\quad \max_{\mathbf{y}\in \mathbb{Z}^{N+1}} \quad &R=\sum_{n\in\mathcal{N}}\left[\frac{y_n}{\frac{1}{r_n}+\frac{(1+d_n)^{y_n-1}}{f_n} }-\lambda y_n c_n\right] \label{eqn:multi1}\\
{\rm{s.t.}} ~\quad &\sum_{n\in\mathcal{N}^+}y_n= K, \label{eqn:multist1}\\
&0\leq y_n \leq M_n,    \quad \forall n\in\mathcal{N}^+,\label{eqn:multist2}
\end{align}
where $\mathbf{y}=(y_0,\cdots, y_N)$ and $c_0=0$, i.e., no cost incurs if a service is hosted by the macro-BS.


Like Problem (P1), Problem (P4) is also an integer nonlinear programming problem, which in general has no efficient method to solve it optimally. Nevertheless, we show in the following that Problem (P4) can be optimally solved, provided that the total number of users $K \leq K^*$, where $K^*$ is the optimal number of the total users that yields the maximum network utility $R$. For the other case that $K>K^*$, the proposed algorithm in the preceding section can be adopted to find a suboptimal solution in an efficient manner.

\vspace{-0.1in}
\subsection{Optimal Load Distribution for $K\leq K^*$}\label{sec:K1}
In this subsection, we develop an optimal relaxation-and-rounding based algorithm to solve Problem (P4), conditioned on $K\leq K^*$. The key idea is to verify that the integer-relaxed Problem (P4) is a convex problem given $K\leq K^*$ and design the optimal rounding method in the sequel.

We first relax the integer $\mathbf{y}$ into real numbers  and solve the relaxed Problem (P4). To proceed, we define the one-sided optimal load of BS $n$ as
\begin{align}\label{eqn:Jn}
J_n = \argmax_{0\leq y_n \leq M_n}\left\{\frac{y_n}{\frac{1}{r_n}+\frac{(1+d_n)^{y_n-1}}{f_n} }-\lambda y_n c_n\right\}.
\end{align}
Clearly, $J_n$ is the amount of load achieving the maximum utility at BS $n$. By taking the first derivative with respect to $y_n$, we derive a general solution of $J_n$ as
\begin{align}\label{eqn:jn}
J_n =\begin{cases}
\min\{J_n^\prime, M_n\}, &\text{if ~} \lambda c_n \leq \left.1\middle/\left[\frac{1}{r_{n}}+\frac{1}{f_{n}(1+d_n)}\right]\right.,\\
0, &\text{otherwise},
\end{cases}
\end{align}
where $J_n^\prime\geq 0$ is the root of the following equation:
\begin{align}\label{eqn:root}
\frac{1}{r_n}+ \frac{(1+d_n)^{y_n-1}}{f_n}\left[1-y_n\ln(1+d_n)\right] = \lambda c_n  \left[\frac{1}{r_n}+\frac{(1+d_n)^{y_n-1}}{f_n}\right]^2.
\end{align}
It can be checked that the LHS of \eqref{eqn:root} is monotonically decreasing while the RHS is monotonically increasing over $y_n\geq 0$. Thus, $J_n^\prime$ can be obtained by the simple bisection search.

Now, we make a key observation of the relaxed Problem (P4) under different values of $K$:
\begin{proposition}\label{pro3}
Define  $K^* \triangleq \sum_{n\in\mathcal{N}^+}J_n$ and $R(K)$ as the optimal objective value of the relaxed Problem (P4) in terms of $K$. The following properties hold:
\begin{itemize}
\item Property 1: If $K =K^*$, the optimal load distribution is $y_n^*=J_n, \forall n\in\mathcal{N}^+$.
\item Property 2: If $K <K^*$,  $y_n^*\leq J_n$ and $R$ is strictly concave in $0\leq y_n\leq J_n, \forall n\in\mathcal{N}^+$.
\item Property 3: If $K>K^*$, $y_n^*\geq J_n, \forall n\in\mathcal{N}^+$.
\item Property 4: $R(K)$ monotonically increases in $0\!\leq K\!\leq K^*$  and $R(K^*)>R(K), \forall K \!> \!K^*$.
\end{itemize}
\end{proposition}
\begin{IEEEproof}
See Appendix B.
\end{IEEEproof}

Proposition \ref{pro3} reveals that $K^*$ is the optimal number of users  that the network can accommodate to achieve the maximum network utility. Seen from Property $4$, when $K\!<\!K^*$, provisioning more users' services into the macro-BS can help increase the network utility, mainly because the resources on each BS are under-utilized (i.e., $y_n^*\leq J_n$ by Property 2) after the optimal JMH. On the contrary, when $K>K^*$, there are too many users hosted by the macro-BS such that each BS is overloaded (i.e., $y_n^*\geq J_n$ by Property 3) even after the optimal JMH. In this case, more small-BSs are needed to increase the network capacity and address the overloaded condition.

Using Property 2 in Proposition \ref{pro3},  for an under-utilized system (i.e., $K\!\leq\!K^*$), we can safely impose constraints $y_n\leq J_n$, $\forall n$, into the relaxed Problem (P4) without loss of optimality: \begin{align}\label{eqn:multi4}
\max_{\mathbf{y}\in \mathbb{R}^{N+1}} \quad  &R=\sum_{n\in\mathcal{N}^+}\left[\frac{y_n}{\frac{1}{r_n}+\frac{(1+d_n)^{y_n-1}}{f_n} }-\lambda y_n c_n\right]\\
{\rm{s.t.}} ~\quad &\sum_{n\in\mathcal{N}^+}y_n= K,  \quad \quad 0\leq y_n\leq J_n, ~ \forall n\in\mathcal{N}^+.\nonumber
\end{align}
With the objective function $R$ being concave over the feasible region of $\mathbf{y}$, Problem \eqref{eqn:multi4} is a convex problem and can be readily solved and the details are omitted here due to space limitation.

After solving Problem \eqref{eqn:multi4}, we next propose a rounding method to recover the optimal integer solution to Problem (P4).
\begin{proposition}\label{pro4}
Any solution $\mathbf{y}^*=(y_0^*, \cdots, y_N^*)\in \mathbb{Z}_+^{N+1}$ to Problem (P4) satisfies
\begin{align}\label{eqn:optimalround}
y_n^*\in \left\{\lfloor y_n^\prime\rfloor, ~\lceil y_n^\prime\rceil \right\},
\end{align}
where $\mathbf{y}^\prime =(y_0^\prime, \cdots, y_N^\prime)\in \mathbb{R}^{N+1}_+$ denotes the unique solution of Problem \eqref{eqn:multi4}.
\end{proposition}
\begin{IEEEproof}
See Appendix C.
\end{IEEEproof}

Thanks to Proposition \ref{pro4}, we can dramatically reduce the range of numerical searching the optimal integer $y_n^*$. Moreover, since the recovered $y_n$ has to satisfy the sum  constraint \eqref{eqn:multist1}, we can further derive the optimal rounding rule as follows.
\begin{proposition}[Optimal Rounding Rule]\label{pro5}
Let $R_n(y_n)\triangleq \frac{y_n}{\frac{1}{r_n}+\frac{(1+d_n)^{y_n-1}}{f_n} }-\lambda y_n c_n$, $\forall n$, and $s\triangleq K-\sum_{n\in\mathcal{N}^+}\lfloor y_n^\prime \rfloor$.
The optimal $\mathbf{y}^*$ that solves Problem (P4) is given by
\begin{align}\label{eqn:rule}
y_n^*=\begin{cases}
\lceil y_n^\prime\rceil, &\text{if~}  R_n(\lceil y_n^\prime\rceil) -R_n(\lfloor y_n^\prime\rfloor) \text{~is one of the $s$ largest},\\
\lfloor y_n^\prime\rfloor, &\text{otherwise}.
\end{cases}
\end{align}
\end{proposition}
\begin{IEEEproof}
Based on  Proposition \ref{pro4} and constraint \eqref{eqn:multist1}, the optimal integer $\mathbf{y}^*$ should meet the combinatorial condition that $s$ of the $y_n^*$'s satisfy $y_n^*=\lceil y_n^\prime\rceil$ and the rest $N+1-s$ satisfy $y_n^*=\lfloor y_n^\prime\rfloor$. Among
all the possible combinations, \eqref{eqn:rule} is the one with the maximum $R$ and thus the optimal solution to Problem (P4), which completes the proof.
\end{IEEEproof}

\vspace{-0.15in}
\subsection{Modified Algorithm for $K>K^*$}\label{sec:K2}
It is worth noting that we can simplify the proposed  Algorithm \ref{alg:2} to suboptimally solve Problem (P4) for any $K$. Recall that Algorithm \ref{alg:2} consists of the optimal sum-of-ratios algorithm for solving the relaxed problem and the suboptimal rounding method for recovering a feasible integer solution. For Problem (P4), the same rounding rule \eqref{eqn:round} can be applied in the integer recovery  while  the  sum-of-ratios algorithm design can be simplified as follows.

We start by transforming the relaxed Problem (P4) into a standard sum-of-ratios problem:
\begin{align}\label{eqn:multi2}
\max_{\mathbf{y} \in \mathcal{G}} ~& \sum_{n\in\mathcal{N}^+}\left[\frac{y_n}{\frac{1}{r_n}+\frac{(1+d_n)^{y_n-1}}{f_n} }+y_n z_n\right]
\end{align}
where $\mathcal{G}\triangleq \{\mathbf{y}\in \mathbb{R}^{N+1}| \sum_{n\in\mathcal{N}^+}y_n= K,  ~0\leq y_n \leq M_n, \forall n\}$ and $z_n = \max_n\{\lambda c_n\}- \lambda c_{n}\geq 0$, $\forall n$, is the modified cost coefficient for reshaping the objective function into a form of sum of non-negative functions.

The sum-of-ratios Problem \eqref{eqn:multi2} can be solved by two-layer optimization. The inner layer is to find the optimal solution to the subtractive-form problem with given auxiliary parameters $(\alpha_n, \beta_n, \gamma_n)$, which is a convex problem expressed as
\begin{align}
\max_{\mathbf{y}\in \mathcal{F}}~\sum_{n\in\mathcal{N}^+} \alpha_n\left[y_n-\beta_n\left(\tfrac{1}{r_n}+\tfrac{(1+d_n)^{y_n-1}}{f_n}\right)\right]+\sum_{n\in\mathcal{N}^+} \left[ y_n z_n-\gamma_n\right] \label{eqn:multi3}.
\end{align}

The subtractive-form Problem \eqref{eqn:multi3} can be easily solved through the following proposition.
\begin{proposition}\label{pro6}
The optimal load distribution that solves Problem  \eqref{eqn:multi3} is
\begin{align}\label{eqn:pro6}
y_n^*=\left[1+\ln \left(\frac{(\alpha_n+z_n -\nu)f_n}{\alpha_n \beta_n \ln(1+d_n)}\right)\right]_0^{M_n},\quad \forall n\in \mathcal{N}^+,
\end{align}
where $[\cdot]_a^b=\max\{a, \min\{\cdot, b\}\}$. $\nu$ satisfying $\sum_{n\in\mathcal{N}^+} y_n^*\!=\!K$ can be obtained by  bisection search.
\end{proposition}

In the outer layer, we use the modified Newton method to find the optimal $(\alpha_n^*, \beta_n^*,  \gamma_n^*)$  like Algorithm \ref{alg:2}. Thus, we omit the detailed description of the sum-of-ratios algorithm for Problem (P4) when $K>K^*$.


\vspace{-0.1in}
\section{Simulation Results}

In this section, we perform simulation to evaluate the performance of our proposed algorithms. We consider $N\!=\!7$ BSs deployed in a square area of $1$ $\text{km}^2$ with a regular hexagonal-lattice layout (see \cite[Fig. 2]{layout}). All the users are randomly distributed within the area at the beginning and their BS associations are initialized using the conventional max-SINR association scheme. We adopt the Random Waypoint Mobility model \cite{RWP} to generate the new user's locations for the considered time slot, with the parameters taken as: the static probability and pause time $p_s=t_p=0$, and the user velocities chosen uniformly at random within the interval $[v_{\text{min}}, v_{\text{max}}]\!=\![0, 5]$ m/s. We set the JMH cost $c_{k,j,n}=W_0+W_k$ if $j\neq n$, and $c_{k,j,n}=0$ otherwise, where $W_0=10^5$ is the handover cost while $W_k$ denotes the VM migration cost, which is  chosen from the set $\{1, 2, 5\}\times 10^5$ according to user $k$'s subscribed service.
Unless mentioned otherwise, the main communication and computation parameters used in the simulations are summarized in Table \ref{tab:1}.
\begin{table}[t]
\linespread{1}
\setlength{\abovecaptionskip}{0.cm}
\setlength{\belowcaptionskip}{-0.5cm}
\centering
\caption{\label{tab:1}System Parameters}
\begin{tabular}{|c|c|}
\hline
Parameter & Value \\
\hline
Number of BSs, $N$   & $7$   \\
\hline
Number of users, $K$   & $60$   \\
\hline
System bandwidth, $B$ & $20$MHz  \\
\hline
Path loss from user to BS & $128.1+37.6\log_{10}l_{[\text{km}]}$ dB \\
\hline
User transmit power, $p_k$   &  $0.1$ W \\
\hline
Expected computation rate, $f_{k,n}$ & $[0.5\times 10^7, 2\times 10^7]$ bits/sec\\
\hline
Degradation factor, $d_{n}$ & $0.25$\\
\hline
Weight of user $k$'s offloading rate, $\omega_k$ & $1$ \\
\hline
Weight of JMH cost, $\lambda$, & $0.5$ \\
\hline
Number of Monte Carlo simulations &$500$\\
\hline
Maximum number of VM, $M_n$ &$45$ \\
\hline
\end{tabular}
\end{table}

\vspace{-0.2in}
\subsection{JMH in a General Multi-cell MEC System}
\begin{table}[t]
\linespread{1}
\setlength{\abovecaptionskip}{0.cm}
\setlength{\belowcaptionskip}{-0.5cm}
\centering
\caption{\label{tab:2}Sum Utility $[\times 10^6]$ v.s. Number of Users}
\begin{tabular}{|c| c| c| c |c |}
\hline
Number of Users & 6 & 8 & 10 & 60 \\
\hline
Upper Bound & 0.814994  & 1.110704 & 1.306949 & 6.408008  \\
\hline
Optimal via Exhaustive Search & 0.814991  & 1.110697 & 1.306941 & --  \\
\hline
Proposed & 0.814991 & 1.110697 & 1.306941 & 6.407563 \\
\hline
\end{tabular}
\end{table}
In Table \ref{tab:2}, we evaluate the sum utility of the proposed Algorithm \ref{alg:2} in comparison with the globally optimal solution by exhaustive search, and the upper-bound result, referring to the optimal solution of Problem (P1$^\prime$). Note that we only provide the performance of the exhaustive search in a small network size due to its exponential complexity. It can be observed that, the performance gap between the upper bound and the exhaustive search does exist. Meanwhile, we can see that the proposed algorithm achieves the optimal performance, indicating that the proposed rounding method in Algorithm $2$ can efficiently recover the optimal integer solutions from the fractional results of the relaxation stage.


Next, we introduce two benchmark schemes for performance comparison: 1) \emph{No migration:} All the users continue the associations with their original BSs;
2) \emph{Radio-oriented migration:} Each user connects to the BS with the highest value of $r_{k,n}-\lambda c_{k,n}$, which represents the traditional BS handover without considering the system dynamics on the computation side.

\begin{figure}[t]
\setlength{\abovecaptionskip}{0.1cm}
\setlength{\belowcaptionskip}{-0.5cm}
\centering
  \begin{minipage}[t]{0.5\linewidth}
    \centering
    \includegraphics[width=0.9\textwidth]{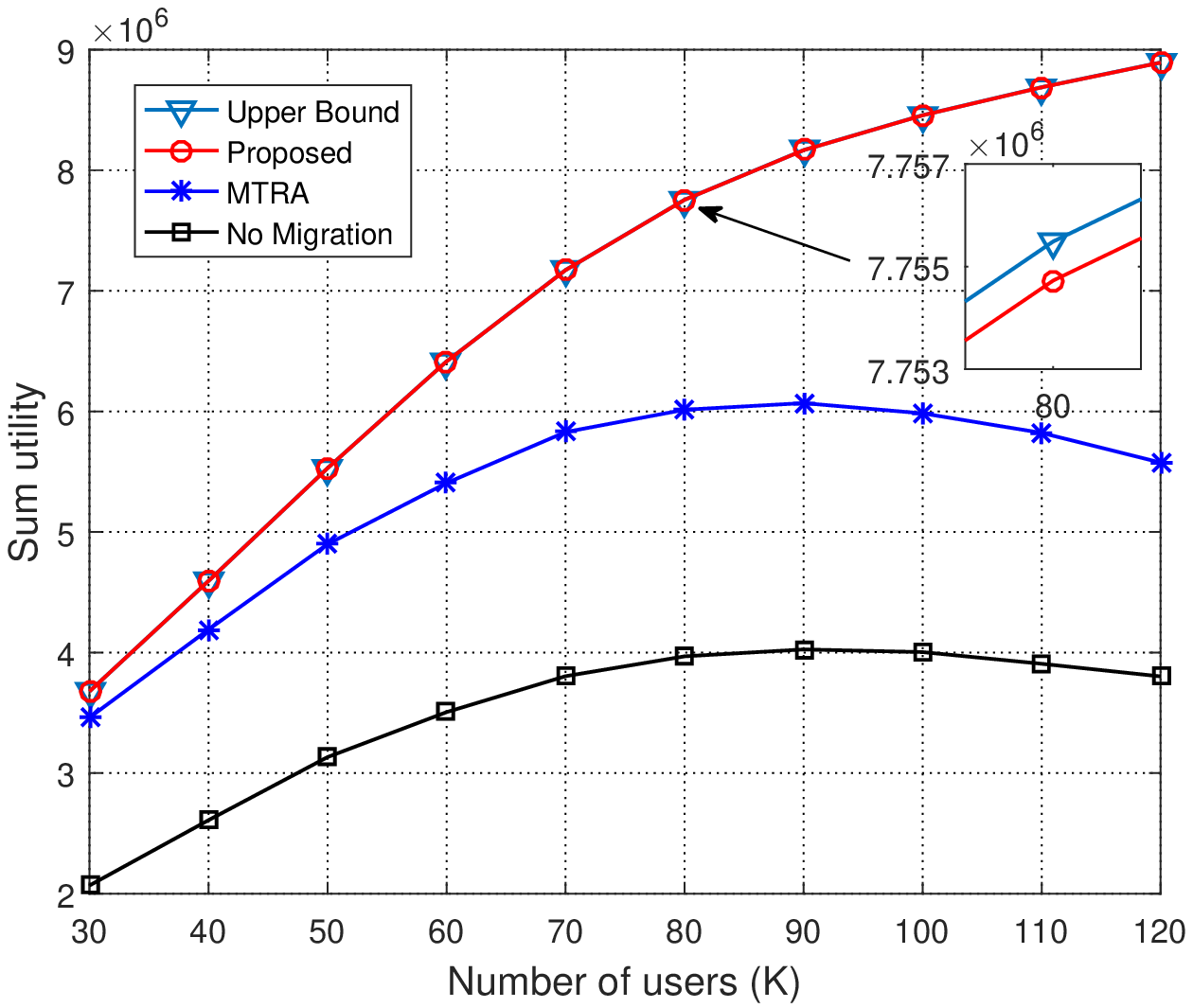}
    \caption{Sum utility vs. $K$.}
    \label{fig:K_vs_utility}
  \end{minipage}%
  \begin{minipage}[t]{0.5\linewidth}
    \centering
    \includegraphics[width=0.9\textwidth]{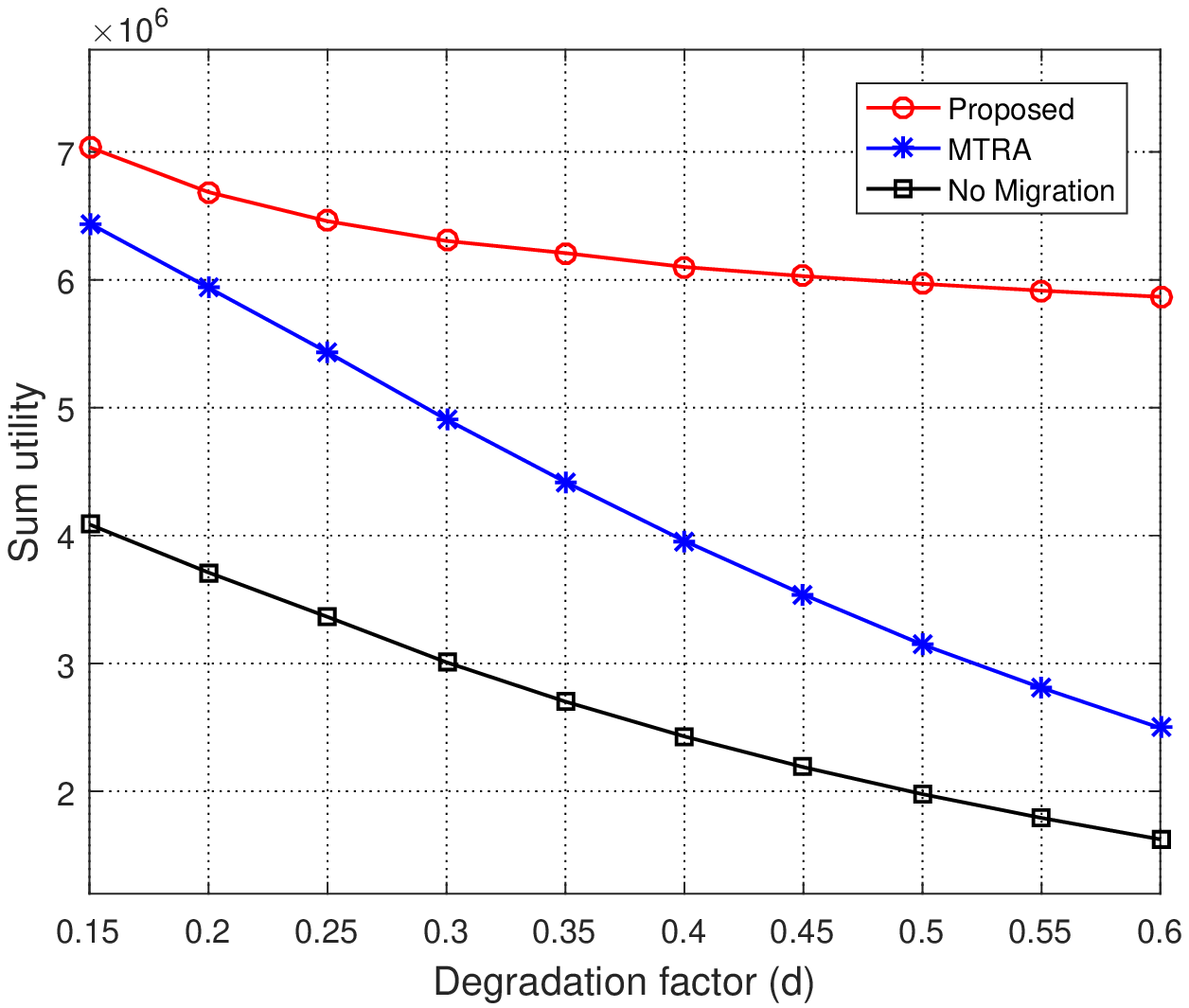}
    \caption{Sum utility vs. $d$.}
    \label{fig:D_vs_utility}
  \end{minipage}
\end{figure}

In Fig. \ref{fig:K_vs_utility}, we compare the sum utility performance of different algorithms versus the number of users $K$. First, we can observe that the performance of the proposed algorithm approaches to the upper bound, indicating  its close-to-optimal performance. The proposed algorithm and radio-oriented migration scheme have large utility gain against the no-migration scheme, since these two schemes jointly manage computation-and-radio resources according to the system dynamics. The radio-oriented migration performs well when $30\!\leq K\!\leq 50$; however it begins to degrade when $K\!\geq\!90$. This is because when $K$ is small, each BS is lightly loaded and wireless channel condition dominates system performance. When $K$ becomes large, the load of each BS becomes varied, leading to notable computation-rate variations among BSs caused by I/O interference. In this case, the radio-oriented migration scheme without considering the computation rate of BSs will suffer severe performance degradation. In contrast, our proposed JMH framework can efficiently mitigate the I/O interference and thus further improve system performance especially when $K$ is large. For instance, when $K\!= \!90$, the proposed algorithm obtains about $34\%$ utility improvement over the radio-oriented migration scheme.

In Fig. \ref{fig:D_vs_utility}, we evaluate the impact of degradation factor on the sum utility performance, where the factor of each BS is set to be identical, i.e., $d_n=d$, $\forall n$. As expected, the proposed algorithm has the slowest descending rate among all the algorithms, showing that our proposed algorithm has the best performance resistance against the I/O interference. We also observe that the performance of radio-oriented migration is close to that of the proposed algorithm when $d$ is small, however, it dramatically decreases when $d$ increases. This is aligned with the discussion in \makebox{Fig. \ref{fig:K_vs_utility}} that the radio-oriented migration performs well when the channel condition is dominant while it has poor performance when the I/O interference becomes a key factor.

\begin{figure}[t]
\setlength{\abovecaptionskip}{0.1cm}
\setlength{\belowcaptionskip}{-0.5cm}
\centering
\subfigcapskip=-10pt
\subfigure[]{\label{sub.v_vs_utility}
    \includegraphics[width=0.45\textwidth]{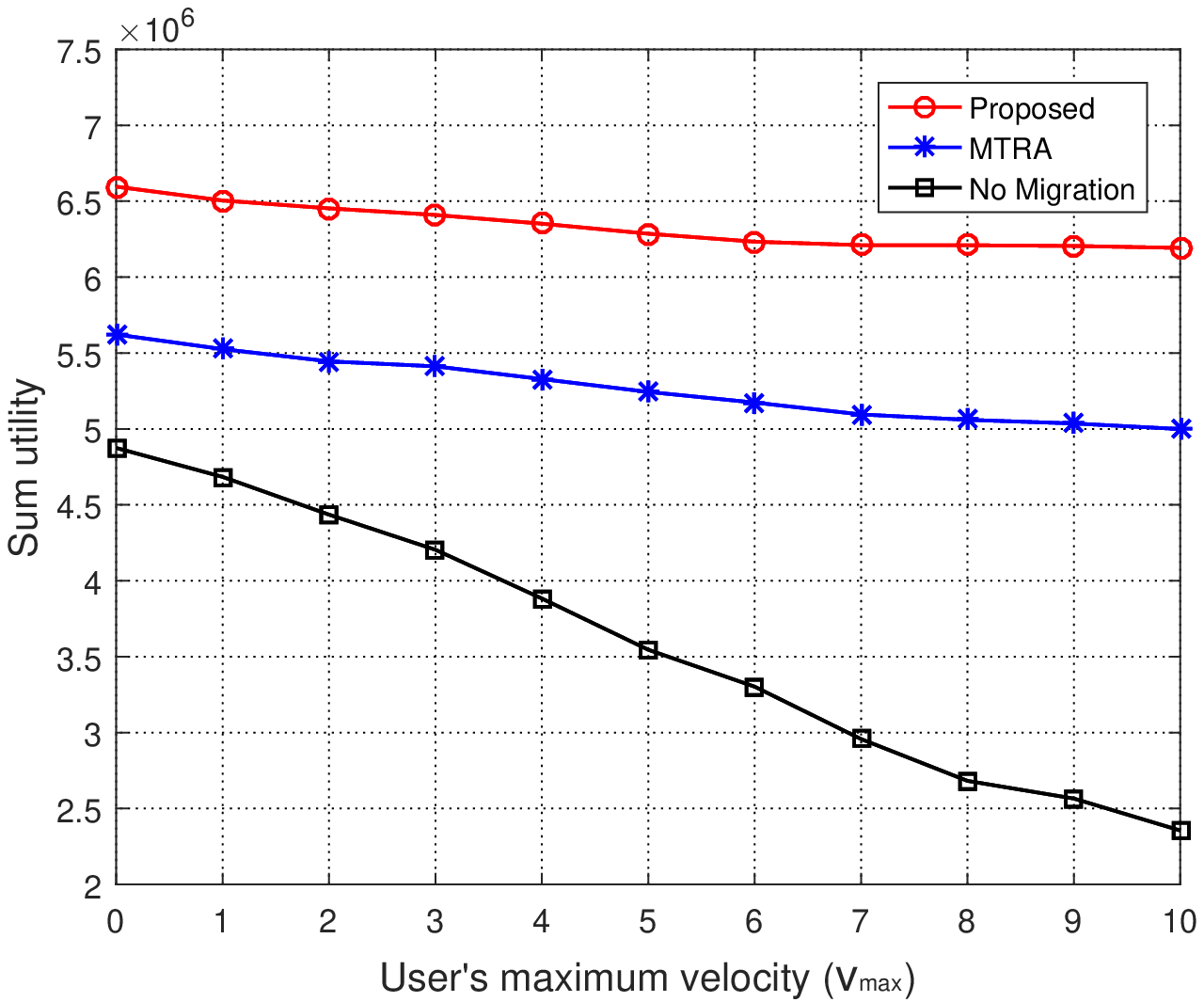}
}
\hspace{-3mm}
\subfigure[]{\label{sub.v_vs_migrated_num}
    \includegraphics[width=0.45\textwidth]{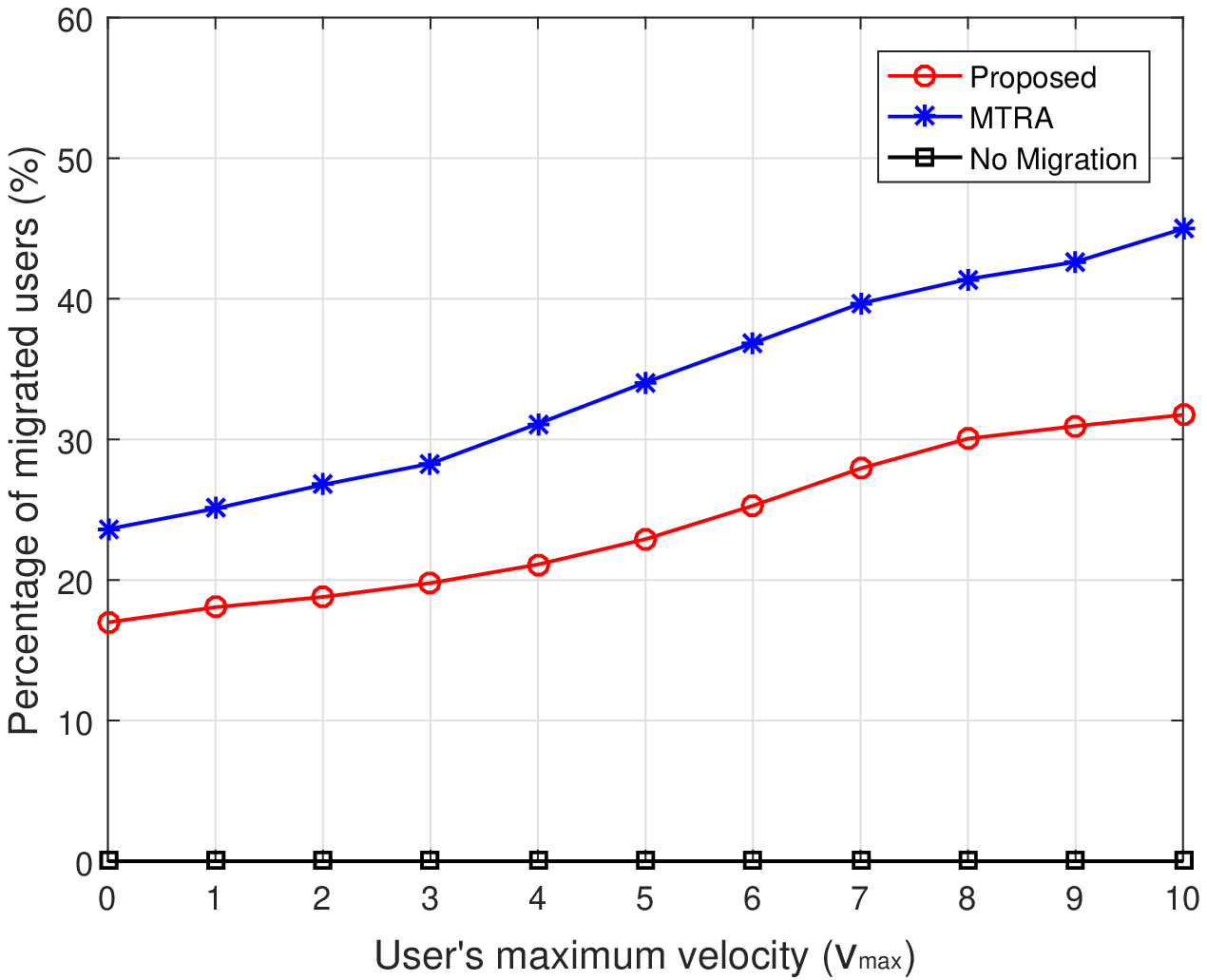}
}
\vspace{-0.1in}
\caption{(a) Sum utility vs. user's maximum velocity $v_{\text{max}}$. ~  (b) Percentage of migrated users vs. $v_{\text{max}}$.}
\end{figure}

Fig. \ref{sub.v_vs_utility} shows the impact of user's mobility on the sum utility, where $v_{\max}$ denotes the user's maximum velocity, with a larger $v_{\text{max}}$ indicating more dramatic location changes and in turn higher channel variations. As expected, the performance of no-migration scheme drastically decreases as $v_{\max}$ increases due to the channel deterioration of the initial BSs. In contrast, the proposed algorithm and radio-oriented migration scheme can efficiently resist the impact of $v_{\max}$, thanks to their flexible user-BS association. On the other hand, when $v_{\max}=0$, i.e., user's location remains static, there are still performance gains achieved by the proposed algorithm and radio-oriented migration compared with no migration. This is because besides user's movement, wireless fading is time-varying, which affects channel condition and thus the JMH policies.

Fig. \ref{sub.v_vs_migrated_num} shows the percentage of migrated users among the total number of users  versus $v_\text{max}$. As we observe, the percentage of migrated users increases with $v_\text{max}$ in both proposed algorithm and radio-oriented migration scheme, which fits our intuition that the user's migration demand grows as the level of mobility increases. Compared with the radio-oriented migration scheme, the proposed algorithm has a lower migration percentage and slower ascending rate against the mobility level; combining with the sum utility behaviors of these two schemes shown in Fig. \ref{sub.v_vs_utility}, these demonstrate that our proposed algorithm can reduce unnecessary migrations and make more accurate migration decisions to improve the sum utility.

\begin{figure}
\setlength{\abovecaptionskip}{0.1cm}
\setlength{\belowcaptionskip}{-0.5cm}
\centering
  \begin{minipage}[t]{0.5\linewidth}
    \centering
    \includegraphics[width=0.9\textwidth]{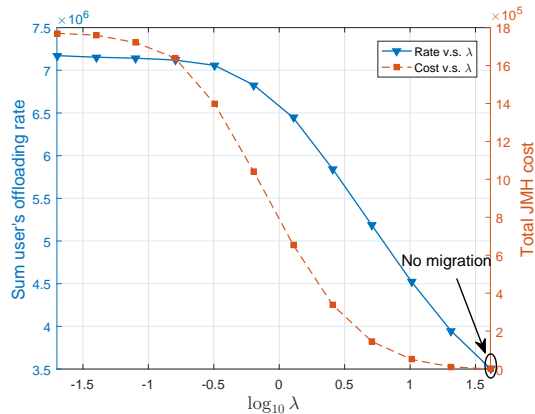}
    \caption{Sum offloading rate/total JMH cost vs. $\lambda$.}
    \label{fig:Rate_cost_tradeoff}
  \end{minipage}%
  \begin{minipage}[t]{0.5\linewidth}
    \centering
    \includegraphics[width=0.9\textwidth]{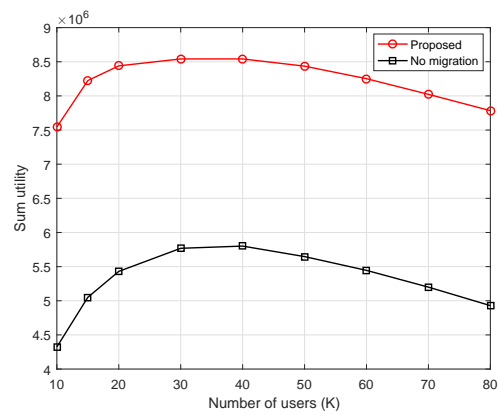}
    \caption{Sum utility vs. $K$ in radio-resource allocation case.}
    \label{fig:K_vs_utility_RA}
  \end{minipage}
\end{figure}

Fig. \ref{fig:Rate_cost_tradeoff} shows the impacts of $\lambda$ on the sum user's offloading rate and total JMH cost of the proposed algorithm. It can be observed that when the price of JMH cost $\lambda$ is small, the proposed algorithm triggers more migrations to improve sum user's offloading rate at the cost of high JMH cost consumption. However, as $\lambda$ increases, the price of doing migration operations increases and our proposed algorithm avoids more worthless migrations (i.e., those with little offloading-rate improvement but at high JMH cost). Therefore, there exists a tradeoff between the sum offloading rate improvement and the JMH cost consumption.  We also observe that setting $\lambda\in[0.8, 5]$ can achieve over $50\%$ offloading-rate improvement compared to the no-migration scheme while maintaining the JMH cost less than half of the maximum JMH cost consumption, which is a desirable interval to balance the performance of these two metrics.

Fig. \ref{fig:K_vs_utility_RA} shows the sum utility versus the number of users $K$ under the radio-resource allocation scenario, where the bandwidth of each BS is set as $B_n\!=\!B/N$ and the user-BS associations are initialized by choosing the BS with the highest value of spectral efficiency $\eta_{k,n}^0$. We can observe that the proposed algorithm has a much larger and more stable performance than the no-migration scheme, thanks to its high spectrum efficiency achieved by radio-resource allocation among users. The performance of the proposed algorithm increases with $K$ when $K$ is small while decreases slowly when $K\geq 40$. This is because when $K$ is small, increasing the number of user at each BS can help leverage the VM-multiplexing gain to further improve the system performance. However, when $K$ is large, the I/O interference becomes the dominant issue of degrading the system performance. In this case, our algorithm can efficiently mitigate the interference so that the system performance decreases at a slower rate than that of the no-migration scheme.

\vspace{-0.15in}
\subsection{Hotspot Mitigation Case}
In this subsection, we turn our attention to the special case of hotspot-mitigation scenario. We consider a macro BS, denoted by BS $0$, with the assistance of $N\!=\!3$ BSs. For BS $0$, we set $[r_0, f_0, d_0]\!=\![5 \text{Mbps}, 5\!\times\! 10^7 \text{bit/s}, 0.25]$. For each BS $n=1,2,3$, we consider a homogenous setting of $[r_n, f_n, c_n, d_n]\!=\![2 \text{Mbps}, 1\!\times\! 10^7 \text{bit/s}, 2\!\times \! 10^5, 0.4]$ for the ease of graphic illustration.

\begin{figure}[t]
\setlength{\abovecaptionskip}{0.1cm}
\setlength{\belowcaptionskip}{-0.5cm}
\centering
\subfigcapskip=-10pt
\subfigure[]{\label{sub.1}
    \includegraphics[width=0.45\textwidth]{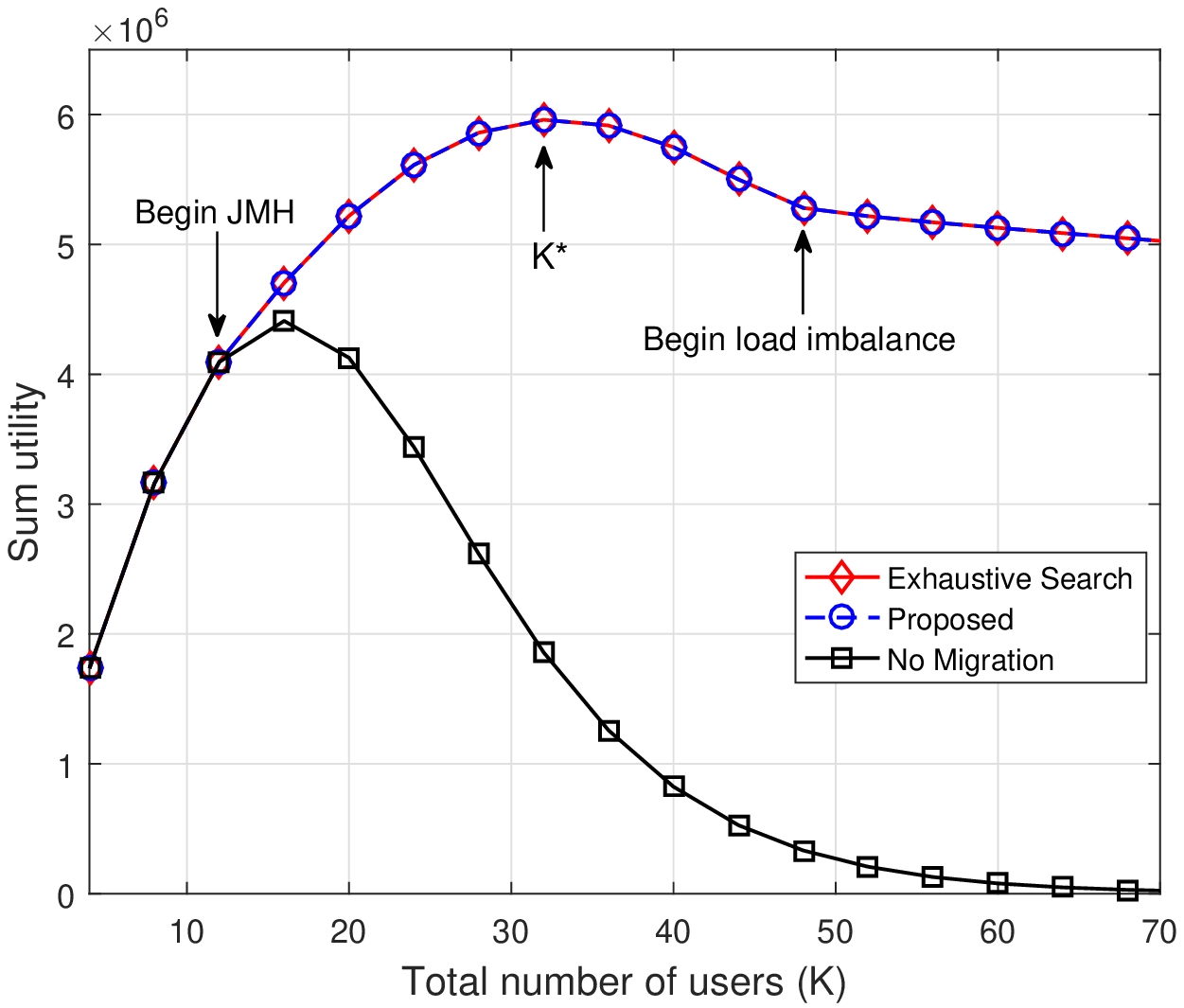}
}
\hspace{-3mm}
\subfigure[]{\label{sub.2}
    \includegraphics[width=0.45\textwidth]{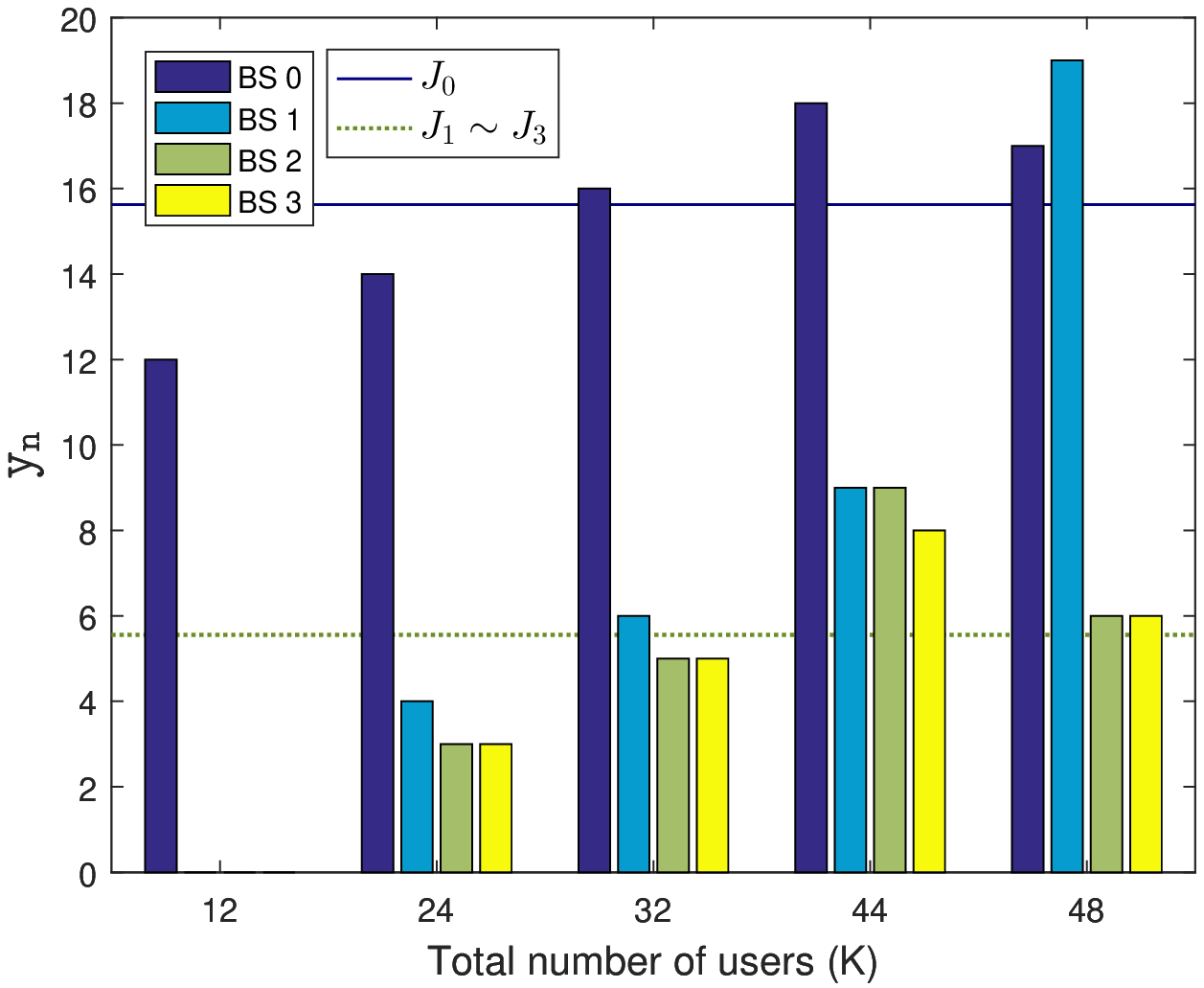}
}
\vspace{-0.1in}
\caption{(a) Sum utility vs. $K$. (b) Load distributions among BSs sampled from Line ``Proposed'' in Fig. \ref{sub.1}.}
\end{figure}

Fig. \ref{sub.1} shows the utility performance of our proposed algorithm versus $K$, where the proposed algorithm includes the relaxation-and-rounding based algorithm to resolve the case $K\!\leq\! K^*$ and the modified Algorithm \ref{alg:2} towards the case $K\!>\!K^*$. For comparison, we also present the optimal performance obtained by exhaustive search and the performance of no-migration scheme mentioned in the preceding section. As can be seen in Fig. \ref{sub.1},  the proposed algorithm can achieve the optimal performance for all $K$, which verifies its optimality behavior when $K\!\leq\! K^*$ and the effectiveness of finding the optimal solution when $K\!>\!K^*$. We also observe that, for both proposed algorithm and no-migration scheme, the utility monotonically increases with $K$ when $K$ is small and it begins to decrease when $K$ exceeds some thresholds due to the computation-rate degradation caused by I/O interference. Nevertheless, compared with the no-migration scheme, our proposed algorithm not only greatly prolongs the utility growth until $K\!>\!K^*$ but also keeps the utility reduction in a much slower rate afterwards.

To illustrate the mechanism behind the optimal JMH scheme, we further analyze the load distribution among BSs shown in Fig. \ref{sub.2}, along with the results shown in Fig. \ref{sub.1}. Specifically, by varying $K$ from $4$ to $70$, the optimal utility goes through the following four stages:
\begin{enumerate}
\item \emph{Stage I ($0\!\leq \!K \!\leq \!12$):} The utility increases with $K$ and no JMH occurs since BS $0$ is still under-utilized. The representative load distribution is $K\!=\!12$ in Fig. \ref{sub.2}.
\item \emph{Stage II ($12\!< \!K \!\leq \! 32$):} Unlike no-migration scheme, the utility of the optimal JMH scheme keeps increasing in this stage thanks to migrating the load to the helper BSs. The load of each BS (i.e., $y_n$) gradually increases as $K$ grows up and all of them are below their one-side optimal load levels [i.e., $J_n$ defined in \eqref{eqn:Jn}]. When $K\!=\!32$, the utility achieves its maximum by the scheme that $y_n \approx J_n, \forall n,$ shown in  $K\!=\!32$. All the observations in this stage verify the results in Proposition \ref{pro3} that when $K \!\leq\! K^*$, $R(K)$ monotonically increases and $y_n^*\!\leq \!J_n, \forall n$; when $K \!=\!K^*$, $R(K)$ achieves its optimum by $y_n^*\!=\!J_n, \forall n$.
\item \emph{Stage III ($32\!<\! K\! \leq \!44$): } The utility begins to decrease. Nevertheless, as shown in $K\!=\!44$, the corresponding optimal JMH scheme is implemented in a load-balance manner, where each BS is lightly overloaded (i.e., slightly above $J_n$) to share the total load, without sacrificing the performance of any one of the BSs.
\item \emph{Stage IV ($K\!>\!44 $):} The utility decreases slowly  in an approximately linear rate. However, contrary to Stage III, it is realized by load imbalance that allocates all the unwanted load into one of the BSs while maintaining the others at their optimal load levels.
\end{enumerate}

To summarize, in the considered hotspot-mitigation scenario, our proposed algorithm can achieve higher utility than no-migration scheme. It performs well especially when the total number of users $K$ is in Stage II or Stage III, which is conducted in efficient resource utilization and load balance among BSs. However, when $K$ is in Stage IV, the system still remains load imbalance after the optimal JMH, implying that there are too many services accommodated at the system and the number of helper BSs is not enough.  In this case, adding more BSs is needed to address the overloaded issue.

\vspace{-0.1in}
\section{Conclusions}
In this paper, we studied the JMH optimization problem in a multi-user multi-cell MEC system, where the I/O interference is considered. We proposed a novel efficient algorithm to solve the combinatorial problem, which achieves the close-to-optimal performance. In addition, we also considered the JMH design for a special  hotspot-mitigation scenario. We obtained the following useful insights for practical multi-user multi-cell/server MEC design: First, communication aspect dominates the system performance when the number of users is small, and computation is the key factor when the number of users is large due to the I/O interference. Second, there exists a threshold on the number of users, such that load balance among BSs can be captured within the threshold while load imbalance happens beyond the threshold.

%

\vspace{-0.1in}
\section*{Appendix}
\appendices
\vspace{-0.15in}
\subsection{Proof of Theorem \ref{theorem1}}
By introducing the auxiliary variables $\bm{\beta}=\{\beta_{k,n}\}$ and $\bm{\gamma}=\{\gamma_{k,n}\}$, Problem (P1$^\prime$) can be equivalently transformed as
\begin{align}
\max_{\substack{(\mathbf{X}, \mathbf{y})\in \mathcal{F}\\ \bm{\beta}, \bm{\gamma}}} \quad & \sum_{k\in\mathcal{K}}\sum_{n\in\mathcal{N}} \beta_{k,n}+\sum_{k\in\mathcal{K}}\sum_{n\in\mathcal{N}}\gamma_{k,n} \label{eqn:A}\\
{\rm{s.t.}} ~~\quad& x_{k,n}\omega_k\geq \beta_{k,n}\left(\tfrac{1}{r_{k,n}}+\tfrac{(1+d_n)^{y_n-1}}{f_{k,n}}\right),   \quad \forall k, n   \label{eqn:A1}\\
&x_{k,n}z_{k,n}\geq\gamma_{k,n}, \qquad~~\qquad\qquad\qquad \forall k, n \label{eqn:A2}
\end{align}
where $\mathcal{F}$ denotes the set of $(\mathbf{X}, \mathbf{y})$ satisfying the convex constraints \eqref{eqn:st21}-\eqref{eqn:st23}. Clearly, the optimal $(\mathbf{X}^*, \mathbf{y}^*, \bm{\beta}^*, \bm{\gamma}^*)$ to Problem \eqref{eqn:A} satisfies the following conditions:
\begin{align}
\beta_{k,n}^*=\frac{x_{k,n}^*\omega_k}{\frac{1}{r_{k,n}}+\frac{(1+d_n)^{y_n^*-1}}{f_{k,n}}}, \quad \quad  \gamma_{k,n}^*= x_{k,n}^*z_{k,n}, \quad \quad  \forall k, n.\label{eqn:3}
\end{align}

Let $\bm{\alpha}=\{\alpha_{k,n}\}$ and $\bm{\nu}=\{\nu_{k,n}\}$ be the Lagrangian multipliers associated with constraints \eqref{eqn:A1} and \eqref{eqn:A2}, respectively. If  $(\mathbf{X}^*, \mathbf{y}^*, \bm{\beta}^*, \bm{\gamma}^*)$ is the optimal solution to Problem \eqref{eqn:A}, there must exist $\bm{\alpha}^*$ and $\bm{\nu}^*$ such that
\begin{align}
\alpha_{k,n}^* = \frac{1}{\frac{1}{r_{k,n}}+\frac{(1+d_n)^{y_n^*-1}}{f_{k,n}}},\quad \quad \nu_{k,n}^* =1, \quad \quad  \forall k, n.\label{eqn:4}
\end{align}
\eqref{eqn:4} is derived from Fritz-John optimality condition. Its detailed  derivation can be found in the proof of Lemma 2.1 in \cite{sum_of_ratios}. We see that the conditions of $\alpha^*_{k,n}$, $\beta^*_{k,n}$ and $\gamma^*_{k,n}$ in \eqref{eqn:3} and \eqref{eqn:4} are correspondingly equivalent to \eqref{eqn:condi1}-\eqref{eqn:condi3}.

On the other hand, it can be checked that the optimal $(\mathbf{X}^*, \mathbf{y}^*)$ of Problem \eqref{eqn:A} satisfies the Karush-Kuhn-Tucker (KKT) conditions for Problem \eqref{eqn:ob3} for given $(\bm{\alpha},\bm{\beta},\bm{\gamma})=(\bm{\alpha}^*,\bm{\beta}^*,\bm{\gamma}^*)$. Here $\bm{\nu}$ is omitted in Problem \eqref{eqn:ob3} since $\nu_{k,n}^*=1$. As Problem \eqref{eqn:ob3} is convex programming for $\bm{\alpha}\succ 0$ and $\bm{\beta}, \bm{\gamma}\succeq0$, the KKT condition is also sufficient optimality condition. Thus, $(\mathbf{X}^*, \mathbf{y}^*)$ is also the solution of Problem \eqref{eqn:ob3} for $(\bm{\alpha},\bm{\beta},\bm{\gamma})=(\bm{\alpha}^*,\bm{\beta}^*,\bm{\gamma}^*)$, which completes the proof.

\vspace{-0.15in}
\subsection{Proof of Proposition \ref{pro10}}

\underline{Proof of Part a)}: To show $\hat{\mathbf{y}}\in \mathcal{Y}$ is equivalent to verifying that $y_n\leq M_n$, $\forall n \in \mathcal{N}$ and the sum constraint $\sum_{n\in\mathcal{N}}y_n=K$ are met by $\hat{\mathbf{y}}$ simultaneously. The first condition is obvious because with $y_n^\prime \leq M_n$ and a non-negative integer $M_n$, $\hat{y}_n\leq \lceil y_n^\prime \rceil \leq M_n$ holds.  For the second one,
first it is easy to prove that $s$ is is an integer among $[0, N-1]$,  and for a given $s \in [0, N-1]$, $\hat{\mathbf{y}}$ by \makebox{rule \eqref{eqn:round}} can meet the sum constraint as long as there exist at least $s$ BSs with a fractional $y_n^\prime$ for taking the ceil operation. Note that a fractional $y_n^\prime$ naturally implies $\lceil y_n^\prime \rceil \leq M_n$ is met. Consider the non-trivial case $s\geq 1$. Since $y_n^\prime-\lfloor y_n^\prime\rfloor<1$, $\mathbf{y}^\prime$ satisfying the sum constraint must contain at least $s+1$ fractional entries to generate $s\geq 1$, i.e., $s+1$ fractional $y_n^\prime$'s exist when $s\geq 1$ and thus the second condition holds. Summarizing above conditions yields the result.

\underline{Proof of Part b)}: We first prove that $\hat{\mathbf{y}}$ by \eqref{eqn:round} is an optimal solution to the following problem:
\begin{align}\label{eqn:rounding_problem}
\quad \min\nolimits_{\mathbf{y}\in \mathbb{Z}^N} ~\|\mathbf{y}-\mathbf{y}^\prime\|_q, ~ \quad {\rm{s.t.}} ~ \sum\nolimits_{n\in\mathcal{N}} y_n =K,
\end{align}
for a given positive integer $K$ and non-negative real numbers $\mathbf{y}^\prime$ with $\sum_{n\in\mathcal{N}} y_n^\prime =K$, and for every $1\leq q \leq \infty$ norm. Let $\mathbf{y}^*=\{y_n^*\}$ be an optimal solution to Problem \eqref{eqn:rounding_problem}. The proof is provided as follows:
\begin{enumerate}
\item \emph{Show $y_n^*\in\{\lfloor y_n^\prime\rfloor, \lceil y_n^\prime\rceil\}, \forall n\in\mathcal{N}$}: This can be proved by contradiction. Suppose that $\mathbf{y}^*$ contains $y_i^*< \lfloor y_i^\prime\rfloor$ for some $i\in \mathcal{N}$ and $y_j^*>\lceil y_j^\prime\rceil$ for some $j\in \mathcal{N}$ with $j\neq i$. Then, we repeatedly construct a better $\overline{\mathbf{y}}=\{\overline{y}_n\}$ by letting
    \begin{align}
    \overline{y}_n= y_n^* \text{~~for~} n\notin \{i,j\}, ~\quad \overline{y}_i=y_i^*+1 \leq \lfloor y_i^\prime\rfloor, \quad \overline{y}_j=y_j^*-1 \geq \lceil y_j^\prime\rceil,
    \end{align}
    such that $\sum \overline{y}_n = \sum y_n^* =K$, and
    \begin{align}
    \|\overline{\mathbf{y}}-\mathbf{y}^\prime\|_q^q &= \sum\nolimits_n \left(\left|y_n^*-y_n^\prime\right|\right)^q+\sum\nolimits_i \left(|y_i^*-y_i^\prime|-1\right)^q + \sum\nolimits_j\left(|y_j^*-y_j^\prime|-1\right)^q \nonumber \\
    &< \sum\nolimits_n \left(\left|y_n^*-y_n^\prime\right|\right)^q+\sum\nolimits_i \left(|y_i^*-y_i^\prime|\right)^q + \sum\nolimits_j\left(|y_j^*-y_j^\prime|\right)^q =  \|\mathbf{y}^*-\mathbf{y}^\prime\|_q^q,
    \end{align}
    until either set $\{i\}$ or set $\{j\}$ is empty. This contradicts to the optimality assumption and thus the above case does not hold.  Suppose that $\mathbf{y}^*$ contains $y_i^*< \lfloor y_i^\prime\rfloor$ for some $i\in \mathcal{N}$ while $y_n^* \in \{\lfloor y_n^\prime\rfloor, \lceil y_n^\prime\rceil\}$ is satisfied for $n \neq i$. For this case, we also can construct a $\overline{\mathbf{y}}$ by letting $\overline{y}_i = y_i^*+1$ and $\overline{y}_k = y_k^*-1$, with $y_k^*=\lceil y_k^\prime\rceil\neq y_k^\prime$, such that $\sum \overline{y}_n =K$ and $\|\overline{\mathbf{y}}-\mathbf{y}^\prime\|_q^q <\|\mathbf{y}^*-\mathbf{y}^\prime\|_q^q$ ( by $\left|\lceil y_k^\prime\rceil-y_k^\prime\right|-\left|\lfloor y_k^\prime\rfloor-y_k^\prime\right|<1$), until set $\{i\}$ is empty. Note that $y_k$ always exists if set $\{i\}\neq \emptyset$ due to $\sum_{n\in\mathcal{N}} y_n^*=K$. Thus, the above case does not hold for optimality. The opposite case $\mathbf{y}^*$ contains $y_j^*>\lceil y_j^\prime\rceil$ for some $j$ is also not optimal, proved by the similar approach. Summarizing the above discussion yields the result.
\item \emph{Show $\hat{\mathbf{y}}$ is the optimal among set $\mathcal{H}\triangleq\{\mathbf{y}\in \mathbb{Z}^N| y_n\in\{\lfloor y_n^\prime\rfloor, \lceil y_n^\prime\rceil\},  \sum_{n\in\mathcal{N}}y_n=K\}$:} As $y_n\in \{\lfloor y_n^\prime\rfloor, \lceil y_n^\prime\rceil\}$ for all $n$,  $\mathbf{y}\in \mathbb{Z}^N$ must contain $s$ entries to be $y_n=\lceil y_n^\prime\rceil$ and the rest to be $y_n=\lfloor y_n^\prime\rfloor$ to preserve $\sum_{n\in\mathcal{N}}y_n=K$. Evidently, $\hat{\mathbf{y}}$ by \makebox{rule \eqref{eqn:round}} can obtain smallest value of $\|\mathbf{y}-\mathbf{y}^\prime\|_q^q$ among all $\mathbf{y}\in \mathcal{H}$ since rounding upwards $s$ entries with the largest $y_n^\prime -\lfloor y_n^\prime\rfloor$ achieves the maximum rounding-error reduction from recovering $y_n=\lfloor y_n^\prime\rfloor$, $\forall n$ to $y_n\in\{\lfloor y_n^\prime\rfloor, \lceil y_n^\prime\rceil\}$ and $\sum_{n\in\mathcal{N}}y_n=K$.
\end{enumerate}

Since $\hat{\mathbf{y}}$ is an optimal solution to Problem \eqref{eqn:rounding_problem} and satisfies $\hat{y}_n\leq M_n$ for all $n$ according to the result of Part i), $\hat{\mathbf{y}}$ is also an optimal solution to Problem \eqref{eqn:rounding_problem} with the additional constraints $y_n\leq M_n$, $\forall n \in \mathcal{N}$, i.e., $\hat{\mathbf{y}}\in\arg\min_{\mathbf{y}\in\mathcal{Y}}\|\mathbf{y}-\mathbf{y}^\prime\|_q$, ending the proof.

\vspace{-0.15in}
\subsection{Proof of Proposition \ref{pro3}}
To prove this proposition, we need the following property of $R$, which can be easily verified by taking the first derivative of $R$ with respect to $y_n$.
\begin{lemma}\label{pro7}
The objective value $R$ is monotonically increasing with $y_n$ in $[0, J_n^\prime]$ and monotonically decreasing when $y_n> J_n^\prime$, for all $n\in\mathcal{N}^+$.
\end{lemma}

\emph{Property 1:} It is due to the facts that i) $y_n=J_n, \forall n\in\mathcal{N}^+$ can achieve the maximal service migration utility at each BS and thus maximize the sum utility $R$;  and ii) $\sum_{n\in\mathcal{N}^+}J_n=K^*$ meets the constraint \eqref{eqn:multist1}.

\emph{Property 2:} Given $K<K^*$, $y_n^*\leq J_n, \forall n\in\mathcal{N}^+$ can be verified by contradiction as follows. Since $K<K^*$, it follows that $\sum_{n\in\mathcal{N}^+}y_n^*<\sum_{ n\in\mathcal{N}^+}J_n$ and there always exists an non-empty subset $\widetilde{\mathcal{N}}\subseteq\mathcal{N}^+$ such that $y_n^*<J_n, \forall n\in \widetilde{\mathcal{N}}$. Suppose that there exist $y_i^*>J_i$ for some $i\in\mathcal{N}^+$. By Lemma \ref{pro7}, we can always find a larger $R$ by decreasing $y_i^*$ and meanwhile increasing some $y_n^*$'s with $n\in\widetilde{\mathcal{N}}$ to rein in the constraint \eqref{eqn:multist1}, which contradicts the definition of the optimal $y_n^*$. Thus, the optimal workload distribution satisfies  $y_n^*\leq J_n, \forall n\in\mathcal{N}^+$.

We first prove the concavity of $R$ in $0\leq y_n\leq J_n^\prime, \forall n\in\mathcal{N}^+$. Take the second derivative of $R$ with respect to $y_n$:
\begin{small}
\begin{align}
\frac{d^2 \!R}{d y_n^2}\!= \!\underbrace{\frac{-2\left(\frac{1}{r_n}\!+\! \frac{(1+d_n)^{y_n\!-\!1}}{f_n}\!\left[1-y_n\ln(1\!+\!d_n)\right]\right)(1\!+\!d_n)^{y_n\!-\!1} \ln(1\!+\!d_n)}{f_n\left[\frac{1}{r_n}+\frac{(1+d_n)^{y_n\!-\!1}}{f_n}\right]^3}}_{A} + \underbrace{\frac{-y_n (1\!+\!d_n)^{y_n\!-\!1} \ln^2(1\!+\!d_n)}{f_n\left[\frac{1}{r_n}+\frac{(1+d_n)^{y_n\!-\!1}}{f_n}\right]^2}}_{B}.
\end{align}
\end{small}
\vspace{-0.2in}

\noindent
As $B\leq 0$ and the denominator of $A$ is positive for $y_n\geq 0$, to prove $\frac{d^2 R}{d y_n^2}\leq 0$ in $0\leq y_n\leq J_n^\prime$, it is sufficient to show the term $g(y_n)\triangleq\frac{1}{r_n}\!+\! \frac{(1+d_n)^{y_n\!-\!1}}{f_n}\left[1-y_n\ln(1\!+\!d_n)\right]$ in $A$ is positive, $\forall y_n \in [0, J_n^\prime]$. Consider the non-trivial case $J_n^\prime>0$. According to the definition of $J_n^\prime$ in \eqref{eqn:root}, we have $g(J_n^\prime)>0$. Also, it it easily proved that $g(y_n)$ monotonically decreases with $y_n\geq0$.  Thus, $g(y_n)\geq g(J_n^\prime)>0, \forall y_n\in[0, J_n^\prime]$ and the strong concavity of $R$ holds in $0\leq y_n\leq J_n^\prime$. Note that $J_n\leq [J_n^\prime]^+$ by \eqref{eqn:jn}. Hence  the strong concavity of $R$ is also valid in $0\leq y_n\leq J_n$.

\emph{Property 3:} Similar to the proof of Property 2 and thus omitted.

\emph{Property 4:} Let $K_1, K_2 \in [0, K^*]$ and $K_1\!<\!K_2$. Define $\{y_n^{(1)}\}_{n\in\mathcal{N}^+}$ as the optimal solution to the integer-relaxed Problem (P4) for given $K_1$. Since $K_1=\sum_{n\in\mathcal{N}^+} y_n^{(1)}<K_2\leq \sum_{n\in\mathcal{N}^+}J_n$ and $y_n^{(1)}\leq J_n, \forall n\in\mathcal{N}^+$ by Property 2, there always exists an increment $\{\delta_n\geq 0\}$ that meets $\sum_{n\in\mathcal{N}^+} (y_n^{(1)}+\delta_n)=K_2$ and $y_n^{(1)}+\delta_n\leq J_n, \forall n\in\mathcal{N}^+$. Then, for any $K_1, K_2$,  we have
\begin{align}\label{eqn:k1k2}
R(K_1)\overset{(a)}{<} \sum_{n\in\mathcal{N}^+}\left[\frac{y_n^{(1)}+\delta_n}{\frac{1}{r_n}+\frac{(1+d_n)^{y_n^{(1)}+\delta_n-1}}{f_n} }-\lambda (y_n^{(1)}+\delta_n) c_n\right] \overset{(b)}{\leq} R(K_2),
\end{align}
where $(a)$ is derived by the monotonically increasing property of $R$ when $y_n\in [0, J_n]$ in Lemma \ref{pro7} and $(b)$ is because $\{y_n^{(1)}+\delta_n\}_{n\in\mathcal{N}^+}$ is a feasible solution to the continuous relaxation of Problem (P4) given $K_2$. Hence, $R(K)$ is monotonically increasing with $K$ in $[0, K^*]$.

As $R(K^*)$ is the optimal objective value of the continuous relaxation of Problem (P4) \textit{without} the constraint \eqref{eqn:multist1} by Lemma \ref{pro7}, $R(K^*)$ is the upper bound of $R(K), \forall K$, ending the proof.

\vspace{-0.15in}
\subsection{Proof of Proposition \ref{pro4}}

The results of this proposition is established by employing the theorems in \cite{optimalrounding}, which is presented as follows for convenience.
\begin{lemma}\label{pro8}
Consider an integer programming problem
\begin{align}\label{eqn:integerprob}
\inf_{\mathbf{m}\in\mathbb{Z}_+^{N}} ~f(\mathbf{m}), \quad \quad {\rm{s.t.}} ~ \sum_{i=1}^{N} m_i=M \in \mathbb{Z}_+
\end{align}
where $f: \mathbb{R}_+^N\rightarrow \mathbb{R}$ is a continuous function. Let $\mathbf{m}^\prime$ be the optimal solution to the relaxation to $\mathbb{R}_+^N$ of Problem \eqref{eqn:integerprob}. If $f$ is directionally strictly-convex with each $m_i$, i.e., $\frac{\partial^2 f}{\partial m_i^2}>0$, for  $i=1,\cdots, N$, then $\mathbf{m}^\prime$ is unique and the optimal $\mathbf{m}^*$ to Problem \eqref{eqn:integerprob} satisfies
\begin{align*}
m_i^*\in\{\lfloor m_i^\prime \rfloor, ~\lceil m_i^\prime\rceil\}, \quad i=1,\cdots, N.
\end{align*}
\end{lemma}

For our Problem (P4) with given $K\leq K^*$, since $R$ is strictly concave with each $y_n$ according to Property 2 in Proposition \ref{pro3}, Lemma \ref{pro8} is applicable, and yields the result \eqref{eqn:optimalround}.

\vspace{-0.1in}
\bibliographystyle{IEEEtran}
\bibliography{link}

\begin{thebibliography}{10}
\providecommand{\url}[1]{#1}
\csname url@samestyle\endcsname
\providecommand{\newblock}{\relax}
\providecommand{\bibinfo}[2]{#2}
\providecommand{\BIBentrySTDinterwordspacing}{\spaceskip=0pt\relax}
\providecommand{\BIBentryALTinterwordstretchfactor}{4}
\providecommand{\BIBentryALTinterwordspacing}{\spaceskip=\fontdimen2\font plus
\BIBentryALTinterwordstretchfactor\fontdimen3\font minus
  \fontdimen4\font\relax}
\providecommand{\BIBforeignlanguage}[2]{{%
\expandafter\ifx\csname l@#1\endcsname\relax
\typeout{** WARNING: IEEEtran.bst: No hyphenation pattern has been}%
\typeout{** loaded for the language `#1'. Using the pattern for}%
\typeout{** the default language instead.}%
\else
\language=\csname l@#1\endcsname
\fi
#2}}
\providecommand{\BIBdecl}{\relax}
\BIBdecl

\bibitem{White_Paper}
{European Telecommunications Standards Institute (ETSI)}, ``Mobile-edge
  {C}omputing-{I}ntroductory technical white paper,'' Sept. 2014.

\bibitem{survey}
Y.~Mao, C.~You, J.~Zhang, K.~Huang, and K.~B. Letaief, ``A survey on mobile
  edge computing: The communication perspective,'' \emph{Commun. Surveys
  Tuts.}, vol.~19, no.~4, pp. 2322--2358, Fourthquarter 2017.

\bibitem{BS_handover}
D.~{Xenakis}, N.~{Passas}, L.~{Merakos}, and C.~{Verikoukis}, ``Mobility
  management for femtocells in {LTE}-advanced: Key aspects and survey of
  handover decision algorithms,'' \emph{Commun. Surveys Tuts.}, vol.~16, no.~1,
  pp. 64--91, Firstquarter 2014.

\bibitem{VMmigration}
F.~{Zhang}, G.~{Liu}, X.~{Fu}, and R.~{Yahyapour}, ``A survey on virtual
  machine migration: Challenges, techniques, and open issues,'' \emph{Commun.
  Surveys Tuts.}, vol.~20, no.~2, pp. 1206--1243, Secondquarter 2018.

\bibitem{singleuser1}
W.~Zhang, Y.~Wen, K.~Guan, D.~Kilper, H.~Luo, and D.~O. Wu, ``Energy-optimal
  mobile cloud computing under stochastic wireless channel,'' vol.~12, no.~9,
  pp. 4569--4581, Sept. 2013.

\bibitem{NOMA}
Y.~{Liu}, ``Exploiting {NOMA} for cooperative edge computing,'' vol.~26, no.~5,
  pp. 99--103, Oct. 2019.

\bibitem{Multiuser1}
C.~You, K.~Huang, H.~Chae, and B.~H. Kim, ``Energy-efficient resource
  allocation for mobile-edge computation offloading,'' vol.~16, no.~3, pp.
  1397--1411, Mar. 2017.

\bibitem{Asynchronous}
C.~{You}, Y.~{Zeng}, R.~{Zhang}, and K.~{Huang}, ``Asynchronous mobile-edge
  computation offloading: Energy-efficient resource management,'' vol.~17,
  no.~11, pp. 7590--7605, Nov. 2018.

\bibitem{multiuser2}
X.~Chen, ``Decentralized computation offloading game for mobile cloud
  computing,'' vol.~26, no.~4, pp. 974--983, Apr. 2015.

\bibitem{multiuser3}
M.~Liu and Y.~Liu, ``Price-based distributed offloading for mobile-edge
  computing with computation capacity constraints,'' \emph{{IEEE} {C}ommun.
  {L}ett.}, vol.~7, no.~3, pp. 420--423, Jun. 2018.

\bibitem{bao}
H.~{Bao} and Y.~{Liu}, ``A two-sided matching approach for distributed edge
  computation offloading,'' in \emph{{IEEE} {ICCC}}, Aug. 2019, pp. 535--540.

\bibitem{RL}
S.~{Wang}, M.~{Chen}, X.~{Liu}, C.~{Yin}, S.~{Cui}, and H.~{Vincent Poor}, ``A
  machine learning approach for task and resource allocation in mobile-edge
  computing-based networks,'' \emph{{IEEE} Internet Things J.}, vol.~8, no.~3,
  pp. 1358--1372, 2021.

\bibitem{multiserver3}
Z.~{Zhou}, P.~{Liu}, J.~{Feng}, Y.~{Zhang}, S.~{Mumtaz}, and J.~{Rodriguez},
  ``Computation resource allocation and task assignment optimization in
  vehicular fog computing: A contract-matching approach,'' vol.~68, no.~4, pp.
  3113--3125, Apr. 2019.

\bibitem{UAV2}
Z.~{Yang}, C.~{Pan}, K.~{Wang}, and M.~{Shikh-Bahaei}, ``Energy efficient
  resource allocation in {UAV}-enabled mobile edge computing networks,''
  vol.~18, no.~9, pp. 4576--4589, 2019.

\bibitem{W}
\BIBentryALTinterwordspacing
M.~Armbrust, A.~Fox, R.~Griffith, A.~D. Joseph, R.~Katz, A.~Konwinski, G.~Lee,
  D.~Patterson, A.~Rabkin, I.~Stoica, and M.~Zaharia, ``A view of cloud
  computing,'' \emph{Commun. ACM}, vol.~53, no.~4, pp. 50--58, Apr. 2010.
  [Online]. Available: \url{https://doi.org/10.1145/1721654.1721672}
\BIBentrySTDinterwordspacing

\bibitem{netIO}
X.~Pu, L.~Liu, Y.~Mei, S.~Sivathanu, Y.~Koh, C.~Pu, and Y.~Cao, ``Who is your
  neighbor: Net {I/O} performance interference in virtualized clouds,''
  \emph{{IEEE} {T}rans. {S}erv. {C}omput}, vol.~6, no.~3, pp. 314--329, Jul.
  2013.

\bibitem{VM1}
X.~Pu, L.~Liu, Y.~Mei, S.~Sivathanu, Y.~Koh, and C.~Pu, ``Understanding
  performance interference of {I/O} workload in virtualized cloud
  environments,'' in \emph{{IEEE} {C}loud}, 2010, pp. 51--58.

\bibitem{our_work}
Z.~{Liang}, Y.~{Liu}, T.~{Lok}, and K.~{Huang}, ``Multiuser computation
  offloading and downloading for edge computing with virtualization,'' vol.~18,
  no.~9, pp. 4298--4311, Sept. 2019.

\bibitem{load_balance}
M.~{Mishra}, A.~{Das}, P.~{Kulkarni}, and A.~{Sahoo}, ``Dynamic resource
  management using virtual machine migrations,'' vol.~50, no.~9, pp. 34--40,
  Sept. 2012.

\bibitem{hot_spot}
T.~Wood, P.~Shenoy, A.~Venkataramani, and M.~Yousif, ``Sandpiper: Black-box and
  gray-box resource management for virtual machines,'' \emph{Computer
  Networks}, vol.~53, no.~17, pp. 2923--2938, 2009.

\bibitem{Migration_policy}
L.~{Gkatzikis} and I.~{Koutsopoulos}, ``Mobiles on cloud nine: Efficient task
  migration policies for cloud computing systems,'' in \emph{{IEEE}
  {CloudNet}}, Oct. 2014, pp. 204--210.

\bibitem{handover_timing}
W.~{Bao}, D.~{Yuan}, Z.~{Yang}, S.~{Wang}, W.~{Li}, B.~B. {Zhou}, and A.~Y.
  {Zomaya}, ``Follow me fog: Toward seamless handover timing schemes in a fog
  computing environment,'' vol.~55, no.~11, pp. 72--78, Nov. 2017.

\bibitem{multiceil2}
Y.~{Sun}, S.~{Zhou}, and J.~{Xu}, ``{EMM}: Energy-aware mobility management for
  mobile edge computing in ultra dense networks,'' vol.~35, no.~11, pp.
  2637--2646, Nov. 2017.

\bibitem{dynamic}
S.~{Wang}, R.~{Urgaonkar}, M.~{Zafer}, T.~{He}, K.~{Chan}, and K.~K. {Leung},
  ``Dynamic service migration in mobile edge computing based on markov decision
  process,'' vol.~27, no.~3, pp. 1272--1288, 2019.

\bibitem{followme}
T.~{Ouyang}, Z.~{Zhou}, and X.~{Chen}, ``Follow me at the edge: Mobility-aware
  dynamic service placement for mobile edge computing,'' vol.~36, no.~10, pp.
  2333--2345, Oct. 2018.

\bibitem{multi_server}
T.~X. {Tran} and D.~{Pompili}, ``Joint task offloading and resource allocation
  for multi-server mobile-edge computing networks,'' vol.~68, no.~1, pp.
  856--868, Jan 2019.

\bibitem{UAV}
C.~{You} and R.~{Zhang}, ``Hybrid offline-online design for {UAV}-enabled data
  harvesting in probabilistic {LoS} channels,'' vol.~19, no.~6, pp. 3753--3768,
  2020.

\bibitem{parallel_computing_mode}
D.~Bruneo, ``A stochastic model to investigate data center performance and
  {Q}o{S} in {I}aa{S} cloud computing systems,'' vol.~25, no.~3, pp. 560--569,
  Mar. 2014.

\bibitem{sum_of_ratios}
\BIBentryALTinterwordspacing
Y.~Jong, ``An efficient global optimization algorithm for nonlinear
  sum-of-ratios problem,'' \emph{Optimization Online}, 2012. [Online].
  Available: \url{http://www.optimization-online.org/DB_FILE/2012/08/3586.pdf}
\BIBentrySTDinterwordspacing

\bibitem{boyd2004convex}
S.~Boyd and L.~Vandenberghe, \emph{Convex optimization}.\hskip 1em plus 0.5em
  minus 0.4em\relax Cambridge university press, 2004.

\bibitem{two_known}
K.~M. Anstreicher and L.~A. Wolsey, ``Two ``well-known'' properties of
  subgradient optimization,'' \emph{Mathematical Programming}, vol. 120, no.~1,
  pp. 213--220, Aug. 2009.

\bibitem{primal_converge}
E.~Gustavsson, M.~Patriksson, and A.-B. Str{\"o}mberg, ``Primal convergence
  from dual subgradient methods for convex optimization,'' \emph{Mathematical
  Programming}, vol. 150, no.~2, pp. 365--390, May 2015.

\bibitem{graph_construct}
Y.~{Liu}, ``Optimal mode selection in {D2D}-enabled multibase station
  systems,'' vol.~20, no.~3, pp. 470--473, Mar. 2016.

\bibitem{hungarian}
H.~W. Kuhn, ``The hungarian method for the assignment problem,'' \emph{Naval
  research logistics quarterly}, vol.~2, no. 1-2, pp. 83--97, 1955.

\bibitem{layout}
W.~{Tang} and S.~{Feng}, ``User selection and power minimization in full-duplex
  cloud radio access networks,'' vol.~67, no.~9, pp. 2426--2438, May 2019.

\bibitem{RWP}
C.~{Bettstetter}, G.~{Resta}, and P.~{Santi}, ``The node distribution of the
  random waypoint mobility model for wireless ad hoc networks,'' vol.~2, no.~3,
  pp. 257--269, Jul. 2003.

\bibitem{optimalrounding}
R.~Cont and M.~Heidari, ``Optimal rounding under integer constraints,''
  \emph{arXiv preprint arXiv:1501.00014}, 2014.

\end{thebibliography}

 \end{document}